\documentclass[a4paper,UKenglish,cleveref, autoref, thm-restate]{lipics-v2021}
\pdfoutput=1 %uncomment to ensure pdflatex processing (mandatatory e.g. to submit to arXiv)
\hideLIPIcs  %uncomment to remove references to LIPIcs series (logo, DOI, ...), e.g. when preparing a pre-final version to be uploaded to arXiv or another public repository

\usepackage{isabelle,isabellesym}

\isabellestyle{it}

\title{Formalising Fisher's Inequality: Formal Linear Algebraic Proof Techniques in Combinatorics}

\titlerunning{Formalising Fisher's Inequality}

\author{Chelsea Edmonds\footnote{Corresponding Author}}{University of Cambridge Department of Computer Science and Technology, UK \and \url{https://www.cst.cam.ac.uk/people/cle47} }{cle47@cam.ac.uk}{https://orcid.org/0000-0002-8559-9133}{Funded jointly by the Cambridge Trust (Cambridge Australia Scholarship) and a Cambridge Department of Computer Science Qualcomm Premium Research Scholarship}%TODO mandatory, please use full name; only 1 author per \author macro; first two parameters are mandatory, other parameters can be empty. Please provide at least the name of the affiliation and the country. The full address is optional. Use additional curly braces to indicate the correct name splitting when the last name consists of multiple name parts.

\author{Lawrence C. Paulson }{University of Cambridge Department of Computer Science and Technology, UK}{lp15@cam.ac.uk}{https://orcid.org/0000-0003-0288-4279]}{Supported by the ERC Advanced Grant ALEXANDRIA (Project GA 742178)}

\authorrunning{C. Edmonds and L.\,C. Paulson}

\Copyright{Chelsea Edmonds and Lawrence C. Paulson} % LIPIcs license is "CC-BY";  http://creativecommons.org/licenses/by/3.0/


\category{} %optional, e.g. invited paper

\relatedversion{} %optional, e.g. full version hosted on arXiv, HAL, or other respository/website
\supplementdetails[subcategory={Formalisation Repository}]{Software}{https://www.isa-afp.org/entries/Fishers_Inequality.html} %linktext, cite, and subcategory are optional

\acknowledgements{Thanks to Wenda Li, for the helpful pointers on working with linear algebra in Isabelle, and Yiannos Stathopoulos, for the suggestions on utilising SErAPIS to navigate results from past formalisations}%optional

\nolinenumbers %uncomment to disable line numbering

\EventEditors{June Andronick and Leonardo de Moura}
\EventNoEds{2}
\EventLongTitle{13th International Conference on Interactive Theorem Proving (ITP 2022)}
\EventShortTitle{ITP 2022}
\EventAcronym{ITP}
\EventYear{2022}
\EventDate{August 7--10, 2022}
\EventLocation{Haifa, Israel}
\EventLogo{}
\SeriesVolume{237}
\ArticleNo{10}
\begin{document}

\maketitle

\begin{abstract}
	The formalisation of mathematics is continuing rapidly, however combinatorics continues to present challenges to formalisation efforts, such as its reliance on techniques from a wide range of other fields in mathematics. This paper presents formal linear algebraic techniques for proofs on incidence structures in Isabelle/HOL, and their application to the first formalisation of Fisher's inequality. In addition to formalising incidence matrices and simple techniques for reasoning on linear algebraic representations, the formalisation focuses on the linear algebra bound and rank arguments. These techniques can easily be adapted for future formalisations in combinatorics, as we demonstrate through further application to proofs of variations on Fisher's inequality.

\end{abstract}

\section{Introduction}
\label{sec:introduction}
The last decade has seen increasing interest in establishing libraries of formalised mathematics, to verify correctness, gain deeper insights into proofs, and benefit from tools such as automation. While there has recently been an increase in combinatorial formalisations, the field remains under-represented in comparison to more traditional areas of mathematics, and presents several challenges to formalisation efforts. This includes the intuitive nature of traditional proofs, discrepancies in definitions, and a reliance on sometimes surprising results and techniques from other fields of mathematics. Furthermore, current formalisations in combinatorics predominantly focus on graph theory and have often been the result of proving major theorems or verifying algorithms rather than building foundational libraries. Gonthier's well-known four colour theorem formalisation in Coq \cite{gonthierFourColourTheorem2008} is one such example.

Many combinatorial structures, such as graphs and designs, are types of incidence set systems. Unlike more traditional fields of mathematics, a significant amount of research into these structures has been driven by their applications in computer science. The formal verification of some of these applications further motivates the need for foundational formalisations in the field. It is common to see repeated techniques in proofs on these structures, drawing on a wide range of other mathematical fields such as linear algebra, probability, and group theory. From a formalisation perspective, this motivates research in two areas: firstly generalising the formalisation of a proof pattern to easily structure proofs and reduce duplication, and secondly exploring how we may be able to suggest the application of such a proof pattern. 

This paper addresses the first point with a focus on linear algebra, aiming to make formalisation of results using common techniques more accessible. Linear algebra is used in the proofs of many foundational results on incidence systems. While more traditional combinatorial proofs have since been found in some cases, in others linear algebra remains the sole known way to prove a theorem, or presents a much cleaner and general proof \cite{godsilToolsLinearAlgebra}\cite{babaiLINEARALGEBRAMETHODS1988}. 

Currently no formalisations of combinatorial proofs using a linear algebraic methodology exist. This paper presents the formalisation of linear algebraic representations of set systems, building on our prior work in design theory \cite{edmondsModularFirstFormalisation2021} and Isabelle/HOL's linear algebra libraries such as matrices in \cite{thiemannFormalizingJordanNormal2016}. We develop alternate formal reasoning techniques for set systems using these representations, and the formalisation of two particularly notable proof methods: the rank argument and linear bound method. These techniques are applied to the first formalisation of Fisher's inequality, a consequential result on the bounds of set systems, which is also credited for the initial development of the linear algebra method in combinatorics. Informally, the generalised version of Fisher's inequality states that given an incidence set system of $n$ points and $m$ distinct subsets with a constant intersection number, $m$ must be less than or equal to $n$. Godsil \cite{godsilToolsLinearAlgebra} labels Fisher's inequality as a \textit{principle}, stating the result is simply \textit{'too important, and too useful, to be termed a theorem'}. 

This paper begins with (2) a summary of necessary mathematical background and related formalisation work, followed by (3) the formalisation of incidence matrices for set systems. In (4) and (5) we discuss formal methodology for the rank and linear bound techniques respectively, and in (6) present the formalisation of a number of variations of Fisher's inequality. We conclude in (7) with a discussion of the formalisation process when drawing on results from multiple fields, and the challenges and advantages of the linear algebra approach compared to traditional combinatorial proof in a formal environment.

\section{Background}
\label{sec:background}

\subsection{Mathematical Background}
\label{ssec:math_background}

\textit{Incidence set systems} are the foundation of many important structures in combinatorics, including graphs, designs, and hypergraphs. We give the design theoretic definition:
\begin{definition}[Incidence Set System]
    An incidence set system $(\mathcal{V}, \mathcal{B})$ is a collection of subsets $\mathcal{B}$ called \emph{blocks} of a finite set $\mathcal{V}$ of \emph{points}. A system is \emph{simple} if $\mathcal{B}$ is a set, i.e. there are no repeated blocks. A \emph{design} is a finite incidence system with no empty blocks. 
\end{definition}

Incidence structures become interesting mathematically by restricting certain properties to impose structural conditions. We give the four most common properties here \cite{colbournHandbookCombinatorialDesigns2007}: 
\begin{romanenumerate}
    \item \textit{Block size} ($k_{B}$): The cardinality of a block in the set system. A design is incomplete if $k_{B} < v$ for all blocks $B \in \mathcal{B}$, where $v = |\mathcal{V}|$.
    \item \textit{Replication number} ($r_x$): the number of blocks point $x$ occurs in.
    \item \textit{Points index} ($\lambda_T$): The number of blocks a $T$ subset of points occurs in.  
    \item \textit{Intersection number}: The number of points two blocks in a design intersect on.
\end{romanenumerate}

Using these properties we can define certain subtypes of incidence systems. For example, we can model a \textit{constant intersect} family of sets by imposing a condition that the intersection number for any two blocks in a design is constant. Similarly, we can model specific types of designs such as a \textit{pairwise balanced design} (PBD), which has a constant points index $\lambda$ for any pair of points, or a \textit{balanced incomplete block design} (BIBD), which is a PBD with added conditions of uniform block size $k$ and incompleteness.

Linear algebraic techniques can initially appear disparate, however there are a number of basic principles that guide their use \cite{godsilToolsLinearAlgebra}\cite{babaiLINEARALGEBRAMETHODS1988}. Firstly, we define the linear algebraic representation of set systems using matrices, a common representation in both mathematics and computer science applications. For incidence systems, we define an \textit{incidence matrix} based on the incidence relation, and its columns as \textit{incidence vectors}. A point $x$ is \textit{incident} with a block $B$ of the design if $x \in B$. Formally an incidence matrix is defined as follows: 

\begin{definition}[Incidence Matrix]
    For a set of points $\{x_1...x_v\}$ and collection of subsets $\{B_1...B_b\}$, $N$ is a $ v \times b$ incidence matrix of the system defined by: 
    \[
        N_{i,j}=\begin{cases} 1 \text{\; if $x_i \in B_j$} \\
            0 \text{\; if $x_i \notin B_j$}\end{cases}
    \] 
\end{definition}

Matrix and vector representations enable the use of numerous results from linear algebra, which Babai and Frankl summarise \cite{babaiLINEARALGEBRAMETHODS1988}. The basic ideas used in both the linear algebra bound and rank arguments are similar, mapping combinatorial objects to vector or matrix representations and using well known theorems on rank, dimension and linear independence to solve problems in extremal combinatorics \cite{juknaExtremalCombinatorics2011}. The incidence relation is one such basic mapping, which is used in the proof of Fisher's inequality. Further detail on the rank argument, linear algebra bound method, and Fisher's inequality is provided as needed in later sections.

\subsection{Related Formalisation Work}

To date, the majority of combinatorics formalisations have been done in Isabelle/HOL, Coq, Lean, and Mizar. Based on a survey of combinatorial results across these systems, Isabelle/HOL (henceforth Isabelle) appears to have the most significant range of general combinatorial results, with the majority in both computer science and mathematics entries within the Isabelle Archive of Formal Proofs (AFP). Isabelle also has significant libraries in linear algebra, making it ideal for this work. 

Aside from numerous graph theory libraries, of which the most extensive is presented by Noschinksi \cite{noschinskiGraphLibraryIsabelle2015}, there are limited formalisations of incidence set systems in Isabelle, including a small matroid library \cite{keinholzMatroids2018}, and our previous work on design theory \cite{edmondsModularFirstFormalisation2021}, which this work extends. Furthermore, no existing formalisation of Fisher's inequality or formalisations of linear algebraic proofs for combinatorics are known to be available in any system.

Isabelle's linear algebra formalisations are scattered across both the main library and AFP, with several concepts having multiple representations. A key and relevant example of this is matrices, for which there are three main formalisations: Harrison's representation \cite{HarrisonMatrices} in the HOL-Analysis library (ported from HOL-Light), Obua's entry on finite matrices based on lists \cite{obuaProvingBoundsReal2005}, and the matrix and vector library developed as part of the Jordan Normal Form (JNF) library. As Thiemann and Yamada address \cite{thiemannFormalizingJordanNormal2016}, this last library provides significant flexibility over past definitions. It removes the type constraints in Harrison's definition where the dimensions of a matrix are modelled by the size of types, while also enabling reasoning at a more abstract level than \cite{obuaProvingBoundsReal2005}. Additionally, the JNF library formalises a number of necessary results on rank, and provides links to earlier vector representations so a large part of Isabelle's existing linear algebra libraries, such as Aransay's work \cite{aransayFormalizationExecutionLinear2014} can still be used.

\subsection{Isabelle and Locales}

Isabelle/HOL is an interactive proof assistant built on higher order logic. In addition to the necessary formalisation background presented above, it has several features which proved critical in this formalisation. Firstly, Isabelle's Isar proof language \cite{DBLP:phd/dnb/Wenzel02a} makes it easier to structure proofs for straightforward application of general techniques, and provides an easily readable syntax. Sledgehammer, Isabelle's automation system, also proved highly useful throughout the formalisation process. 

This formalisation builds on prior work \cite{edmondsModularFirstFormalisation2021} which emphasises use of locales, Isabelle's module system \cite{ballarinLocalesLocaleExpressions2004}, to establish a flexible and easily extendable mathematical structure hierarchy. Locales are an important element of the Isar proof language, providing persistent contexts which can be used across numerous theories drawing on similar structures. While not a new feature, first introduced in their current form in 2004, their full power for formalising mathematics has only recently been realised. 

In the simplest form, a locale declaration introduces parameters (with a specified type) and assumptions. Once defined, a locale can be extended with definitions, notation, and theorems within its context. Locale expressions were designed to support multiple inheritance diamonds. Existing locales can be combined to create a new locale, and extended by adding new parameters and assumptions. The locale hierarchy can easily be transformed using the \textbf{sublocale} command, which is used to show indirect inheritance between two separately specified locales. It is also possible to instantiate locale parameters and instances through locale expressions and interpretations, in both proof and theory contexts.

\section{Incidence Matrices}
\label{sec:inc_matrices}

The foundation of any linear algebraic proof in combinatorics is the matrix or vector representation. This section covers the formalisation of incidence matrices and basic proof techniques for formal reasoning. We note that the techniques used in this section would be straightforward to adapt to formalise other linear algebraic representations, such as adjacency matrices on graphs.

\subsection{Ordering Incidence Systems}

The first challenge a matrix representation presents is that it is inherently ordered. As can be seen from definition (2), we (arbitrarily) label the blocks and points of a system with numbers so that we are able to refer to a certain point as a row index, and block as a column index. This is similarly required in Isabelle, as the matrix formalisation indexes rows and columns using natural numbers. As such, the existing incidence system formalisation must be adapted to be able to arbitrarily impose an ordering on the block collection and point set. 

While a bijective index function would be suitable for points, bijections on multisets are less well defined and no existing formalisation exists. Instead, we created an alternate representation of incidence systems using lists. This method enables us to utilise the large library of results on lists and translates easily to vector representations. Additionally, it has the advantage of being the representation used in several common programming libraries for designs, such as the GAP library \cite{soicherDesignGAPManual}. We define the \textit{ordered-incidence-system} locale below: 

\medskip\isacommand{locale}\isamarkupfalse%
\ ordered{\isacharunderscore}{\kern0pt}incidence{\isacharunderscore}{\kern0pt}system\ {\isacharequal}{\kern0pt}\isanewline
\ \ \isakeyword{fixes}\ {\isasymV}s\ {\isacharcolon}{\kern0pt}{\isacharcolon}{\kern0pt}\ {\isachardoublequoteopen}{\isacharprime}{\kern0pt}a\ list{\isachardoublequoteclose}\ \isakeyword{and}\ {\isasymB}s\ {\isacharcolon}{\kern0pt}{\isacharcolon}{\kern0pt}\ {\isachardoublequoteopen}{\isacharprime}{\kern0pt}a\ set\ list{\isachardoublequoteclose}\isanewline
\ \ \isakeyword{assumes}\ wf{\isacharunderscore}{\kern0pt}list{\isacharcolon}{\kern0pt}\ {\isachardoublequoteopen}b\ {\isasymin}{\isacharhash}{\kern0pt}\ {\isacharparenleft}{\kern0pt}mset\ {\isasymB}s{\isacharparenright}{\kern0pt}\ {\isasymLongrightarrow}\ b\ {\isasymsubseteq}\ set\ {\isasymV}s{\isachardoublequoteclose}
\ \ \isakeyword{and}\ distinct{\isacharcolon}{\kern0pt}\ {\isachardoublequoteopen}distinct\ {\isasymV}s{\isachardoublequoteclose}\medskip

The flexibility of Isabelle's locale mechanism and our previous set theory formalisation, makes it straightforward to prove this locale is a type of finite incidence system through a sublocale declaration: 

\medskip\isacommand{sublocale}\isamarkupfalse%
\ ordered{\isacharunderscore}{\kern0pt}incidence{\isacharunderscore}{\kern0pt}system\ {\isasymsubseteq}\ finite{\isacharunderscore}{\kern0pt}incidence{\isacharunderscore}{\kern0pt}system\ {\isachardoublequoteopen}set\ {\isasymV}s{\isachardoublequoteclose}\ {\isachardoublequoteopen}mset\ {\isasymB}s{\isachardoublequoteclose}\medskip

This avoids the need for a single locale to contain unnecessary information, such as both the ordered lists and original sets. We additionally use the permutations of set and multiset functions to define an alternate introduction rule for an ordered incidence system. The support of multiple inheritance for locales makes it easy to combine this new locale with various existing sub types of incidence structures and in turn create new ordered contexts for specific properties. Within the \textit{ordered-incidence-system} locale, we provide a range of base lemmas to be able to reason on mappings between the set and list based representations, as well as valid indexes in the orderings. 

\subsection{Constant Intersect Designs}
\label{sec:design_theory}

To enable the formalisation of Fisher's inequality in its multiple forms, a number of extensions are required to the existing design theory library \cite{edmondsModularFirstFormalisation2021}. In particular, constant intersect set systems are crucial to reasoning for the non-uniform Fisher's inequality. While the intersection number was previously defined, limited properties were proven. Using the same approach to building the original design theory hierarchy, we define a new locale to reason on incidence systems with a constant intersection number:

\medskip\isacommand{locale}\isamarkupfalse%
\ const{\isacharunderscore}{\kern0pt}intersect{\isacharunderscore}{\kern0pt}design\ {\isacharequal}{\kern0pt}\ proper{\isacharunderscore}{\kern0pt}design\ {\isacharplus}{\kern0pt}\ \isanewline
\ \ \isakeyword{fixes}\ {\isasymm}\ {\isacharcolon}{\kern0pt}{\isacharcolon}{\kern0pt}\ nat\isanewline
\ \ \isakeyword{assumes}\ const{\isacharunderscore}{\kern0pt}intersect{\isacharcolon}{\kern0pt}\ {\isachardoublequoteopen}b{\isadigit{1}}\ {\isasymin}{\isacharhash}{\kern0pt}\ {\isasymB}\ {\isasymLongrightarrow}\ b{\isadigit{2}}\ {\isasymin}{\isacharhash}{\kern0pt}\ {\isacharparenleft}{\kern0pt}{\isasymB}\ {\isacharminus}{\kern0pt}\ {\isacharbraceleft}{\kern0pt}{\isacharhash}{\kern0pt}b{\isadigit{1}}{\isacharhash}{\kern0pt}{\isacharbraceright}{\kern0pt}{\isacharparenright}{\kern0pt}\ {\isasymLongrightarrow}\ b{\isadigit{1}}\ {\isacharbar}{\kern0pt}{\isasyminter}{\isacharbar}{\kern0pt}\ b{\isadigit{2}}\ {\isacharequal}{\kern0pt}\ {\isasymm}{\isachardoublequoteclose}\medskip

\subsection{Incidence Matrix Construction}
Incidence matrices are relatively straightforward to construct given a point and block listing. To enable reasoning on the equality between a set system and an arbitrary incidence matrix, we define the construction of an incidence matrix outside a designated locale:

\medskip\isacommand{definition}\isamarkupfalse%
\ inc{\isacharunderscore}{\kern0pt}mat{\isacharunderscore}{\kern0pt}of\ {\isacharcolon}{\kern0pt}{\isacharcolon}{\kern0pt}\ {\isachardoublequoteopen}{\isacharprime}{\kern0pt}a\ list\ {\isasymRightarrow}\ {\isacharprime}{\kern0pt}a\ set\ list\ {\isasymRightarrow}\ {\isacharparenleft}{\kern0pt}{\isacharprime}{\kern0pt}b\ {\isacharcolon}{\kern0pt}{\isacharcolon}{\kern0pt}\ {\isacharbraceleft}{\kern0pt}ring{\isacharunderscore}{\kern0pt}{\isadigit{1}}{\isacharbraceright}{\kern0pt}{\isacharparenright}{\kern0pt}\ mat{\isachardoublequoteclose}\ \isakeyword{where}\isanewline
{\isachardoublequoteopen}inc{\isacharunderscore}{\kern0pt}mat{\isacharunderscore}{\kern0pt}of\ Vs\ Bs\ {\isasymequiv}\ mat\ {\isacharparenleft}{\kern0pt}length\ Vs{\isacharparenright}{\kern0pt}\ {\isacharparenleft}{\kern0pt}length\ Bs{\isacharparenright}{\kern0pt}\ {\isacharparenleft}{\kern0pt}{\isasymlambda}\ {\isacharparenleft}{\kern0pt}i{\isacharcomma}{\kern0pt}j{\isacharparenright}{\kern0pt}\ {\isachardot}{\kern0pt}\ if\ Vs\ {\isacharbang}{\kern0pt}\ i\ {\isasymin}\ Bs\ {\isacharbang}{\kern0pt}\ j\ then\ {\isadigit{1}}\ else\ {\isadigit{0}}{\isacharparenright}{\kern0pt}{\isachardoublequoteclose}%
\medskip

This produces a 0-1 matrix which we can define numerous base properties for. Note that we use the \textit{ring-1} type for the matrix elements. This type's properties are sufficient for basic reasoning lemmas and calculations on incidence matrices, and enables easy translations to a number of common field types such as $\mathbb{R}$ and $(\mathbb{Z}/2\mathbb{Z})$ when utilising vector space concepts in later parts of the formalisation. Additionally, we define a similar function to produce an incidence vector for a singular block, and prove lemmas on the relationship between these definitions. Within the \textit{ordered-incidence-system} locale, we use $N$ to notate its incidence matrix which is defined as an integer matrix using this function.

\subsection{Incidence Matrix Properties}

We define a 0-1 matrix context through an assumption on matrix elements for any matrix with elements of type \textit{zero-neq-one}. 

\medskip\isacommand{locale}\isamarkupfalse%
\ zero{\isacharunderscore}{\kern0pt}one{\isacharunderscore}{\kern0pt}matrix\ {\isacharequal}{\kern0pt}\ \isanewline
\ \ \isakeyword{fixes}\ matrix\ {\isacharcolon}{\kern0pt}{\isacharcolon}{\kern0pt}\ {\isachardoublequoteopen}{\isacharprime}{\kern0pt}b\ {\isacharcolon}{\kern0pt}{\isacharcolon}{\kern0pt}\ {\isacharbraceleft}{\kern0pt}zero{\isacharunderscore}{\kern0pt}neq{\isacharunderscore}{\kern0pt}one{\isacharbraceright}{\kern0pt}\ mat{\isachardoublequoteclose}\ {\isacharparenleft}{\kern0pt}{\isachardoublequoteopen}M{\isachardoublequoteclose}{\isacharparenright}{\kern0pt}\isanewline
\ \ \isakeyword{assumes}\ elems{\isadigit{0}}{\isadigit{1}}{\isacharcolon}{\kern0pt}\ {\isachardoublequoteopen}elements{\isacharunderscore}{\kern0pt}mat\ M\ {\isasymsubseteq}\ {\isacharbraceleft}{\kern0pt}{\isadigit{0}}{\isacharcomma}{\kern0pt}\ {\isadigit{1}}{\isacharbraceright}{\kern0pt}{\isachardoublequoteclose}\medskip

The \textit{inc-mat-of} definition clearly produces a matrix satisfying this locale's assumptions. Within this environment, we are able to prove a number of basic properties on 0-1 matrices which are useful for reasoning on incidence matrices. Additionally, we define a way to map a given incidence vector $v$ back to a block, assuming the point set is simply $\{0..<dim(v)\}$. 

It is straightforward to prove this mapping results in an incidence system: given any 0-1 matrix of dimension $(v \times b)$, we can construct a system with $v$ points, and $b$ subsets of those points. This proof is possible due to reasoning on 0-1 matrices and general matrix properties being declared outside of the \textit{ordered-incidence-system} locale context. Additionally, it is possible to easily extend this locale to further restrict the type of the matrix elements as needed, such as a 0-1 integer matrix context.

Incidence matrices present an alternate method for reasoning on key properties of a design. For each key set theoretic property, we provide a matrix definition and establish equivalence outside of a locale context. The replication number definition is given below as an example, where the number of blocks a point occurs in is the number of ones in a row.

\medskip\isacommand{definition}\isamarkupfalse%
\ mat{\isacharunderscore}{\kern0pt}rep{\isacharunderscore}{\kern0pt}num\ {\isacharcolon}{\kern0pt}{\isacharcolon}{\kern0pt}\ {\isachardoublequoteopen}{\isacharparenleft}{\kern0pt}{\isacharprime}{\kern0pt}a\ {\isacharcolon}{\kern0pt}{\isacharcolon}{\kern0pt}\ {\isacharbraceleft}{\kern0pt}zero{\isacharunderscore}{\kern0pt}neq{\isacharunderscore}{\kern0pt}one{\isacharbraceright}{\kern0pt}{\isacharparenright}{\kern0pt}\ mat\ {\isasymRightarrow}\ nat\ {\isasymRightarrow}\ nat{\isachardoublequoteclose}\ \isakeyword{where}\isanewline
{\isachardoublequoteopen}mat{\isacharunderscore}{\kern0pt}rep{\isacharunderscore}{\kern0pt}num\ M\ i\ {\isasymequiv}\ count{\isacharunderscore}{\kern0pt}vec\ {\isacharparenleft}{\kern0pt}row\ M\ i{\isacharparenright}{\kern0pt}\ {\isadigit{1}}{\isachardoublequoteclose}\medskip

We can define similar equivalences for balance, intersection and uniformity properties. Note a number of extensions to the vector, matrix, and multiset libraries occurred during this process, included in the final formalisation. These properties enable us to reason on the balance and uniformity conditions of specific types of systems using the incidence matrix in later proofs. For example, the incidence matrix for a block design with uniform block size $k$ will have $k$ ones in each column. 

\subsection{Incidence Matrices for Simple Proofs}
We can now utilise the properties above to begin proving simple results on incidence systems through these alternate representations. Drawing on introductory results from Brualdi and Ryser \cite{brualdiCombinatorialMatrixTheory1991}, one such proof is on the \textit{complement} of a design. The complement of a design $(\mathcal{V}, \mathcal{B})$ has the same point set, but takes the block complement $\mathcal{V} - B$ for each $B \in \mathcal{B}$. We prove a design complement's incidence matrix simply flips all ones and zeros. 

\medskip\isacommand{lemma}\isamarkupfalse% 
\ ordered{\isacharunderscore}{\kern0pt}complement{\isacharunderscore}{\kern0pt}mat{\isacharunderscore}{\kern0pt}map{\isacharcolon}{\kern0pt}\ \isanewline {\isachardoublequoteopen}ordered{\isacharunderscore}{\kern0pt}comp{\isachardot}{\kern0pt}N\ {\isacharequal}{\kern0pt}\ map{\isacharunderscore}{\kern0pt}mat\ {\isacharparenleft}{\kern0pt}{\isasymlambda}x{\isachardot}{\kern0pt}\ if\ x\ {\isacharequal}{\kern0pt}\ {\isadigit{1}}\ then\ {\isadigit{0}}\ else\ {\isadigit{1}}{\isacharparenright}{\kern0pt}\ N{\isachardoublequoteclose}\medskip

Another important concept is that of isomorphisms on designs. It is straightforward to see that if two designs have the same incidence matrix they must be isomorphic, and similarly that two isomorphic designs must have an ordering that produces the same incidence matrix. We prove this formally by establishing a bijective function using the index function on the ordered lists of points. Lastly, Stinson \cite{stinsonCombinatorialDesignsConstructions2004}, presents a number of results on the existence of certain types of designs based on matrix elements, such as the existence of a regular PBD. 

\begin{theorem}
    Let $N$ be a $v \times b$ 0-1 matrix. Then $N$ is the incidence matrix of a regular pairwise balanced design having $v$ points and $b$ blocks if and only if there exist positive integers $r$ and $\lambda$ such that $NN^T = \lambda J_v + (r - \lambda)I_v$.
\end{theorem}

One half of this theorem is formalised in the 0-1 matrix (integer restricted) locale, the other in the regular pairwise balanced locale, which we give below.

\medskip\isacommand{lemma}\isamarkupfalse%
\ rpbd{\isacharunderscore}{\kern0pt}incidence{\isacharunderscore}{\kern0pt}matrix{\isacharunderscore}{\kern0pt}cond{\isacharcolon}{\kern0pt}\ {\isachardoublequoteopen}N\ {\isacharasterisk}{\kern0pt}\ {\isacharparenleft}{\kern0pt}N\isactrlsup T{\isacharparenright}{\kern0pt}\ {\isacharequal}{\kern0pt}\ {\isasymLambda}\ {\isasymcdot}\isactrlsub m\ {\isacharparenleft}{\kern0pt}J\isactrlsub m\ {\isasymv}{\isacharparenright}{\kern0pt}\ {\isacharplus}{\kern0pt}\ {\isacharparenleft}{\kern0pt}{\isasymr}\ {\isacharminus}{\kern0pt}\ {\isasymLambda}{\isacharparenright}{\kern0pt}\ {\isasymcdot}\isactrlsub m\ {\isacharparenleft}{\kern0pt}{\isadigit{1}}\isactrlsub m\ {\isasymv}{\isacharparenright}{\kern0pt}{\isachardoublequoteclose}\medskip

\subsection{Dual Systems}
Dual systems are a crucial concept in design theory and for combinatorial structures more generally. Intuitively a dual swaps the point and block sets of a system, so each block represents a point in the design, and each point represents a block. The Handbook of Combinatorial Designs \cite{colbournHandbookCombinatorialDesigns2007} defines the \textit{dual set system} for $(\mathcal{V}, \mathcal{B})$ as the system $(\mathcal{B}, \mathcal{V})$, where the point $B$ is in a block representing $x$ if and only if $x \in B$ in the original. This definition, while common, is also ambiguous. It doesn't clearly allow for repeated blocks to be distinct points in the dual, which is necessary for properties of a dual to make sense. The ability to impose an arbitrary numbering on the points and blocks enables us to refine the definition to explicitly allow this, which is presented below:

\begin{definition}[Dual System]
    Let $\{x_1, \dots x_v\}$ be the points of a set system with $\{B_1, \dots, B_b\}$ as the blocks. The dual of the system has points $\mathcal{V}^* = \{1,\dots, b\}$, and blocks $\mathcal{B}^* = \{B^*_1, \dots, B^*_v\}$, where a point $y \in V^*$ is in $B^*_n$ for some $1 \leq n \leq v$ if and only if $x_n \in B_y$.
\end{definition}

Using definition 4, we formalise dual designs in the \textit{ordered-incidence-system} locale. 

\medskip\isacommand{definition}\isamarkupfalse%
\ dual{\isacharunderscore}{\kern0pt}blocks{\isacharunderscore}{\kern0pt}ordered\ {\isacharcolon}{\kern0pt}{\isacharcolon}{\kern0pt}\ {\isachardoublequoteopen}nat\ set\ list{\isachardoublequoteclose}\ {\isacharparenleft}{\kern0pt}{\isachardoublequoteopen}{\isasymB}s{\isacharasterisk}{\kern0pt}{\isachardoublequoteclose}{\isacharparenright}{\kern0pt}\ \isakeyword{where}\isanewline
{\isachardoublequoteopen}dual{\isacharunderscore}{\kern0pt}blocks{\isacharunderscore}{\kern0pt}ordered\ {\isasymequiv}\ map\ {\isacharparenleft}{\kern0pt}{\isasymlambda}\ x\ {\isachardot}{\kern0pt}\ {\isacharbraceleft}{\kern0pt}y\ {\isachardot}{\kern0pt}\ y\ {\isacharless}{\kern0pt}\ length\ {\isasymB}s\ {\isasymand}\ x\ {\isasymin}\ {\isasymB}s\ {\isacharbang}{\kern0pt}\ y{\isacharbraceright}{\kern0pt}{\isacharparenright}{\kern0pt}\ {\isasymV}s{\isachardoublequoteclose}\isanewline
\isacommand{interpretation}\isamarkupfalse%
\ ordered{\isacharunderscore}{\kern0pt}dual{\isacharunderscore}{\kern0pt}sys{\isacharcolon}{\kern0pt}\ ordered{\isacharunderscore}{\kern0pt}incidence{\isacharunderscore}{\kern0pt}system\ {\isachardoublequoteopen}{\isacharbrackleft}{\kern0pt}{\isadigit{0}}{\isachardot}{\kern0pt}{\isachardot}{\kern0pt}{\isacharless}{\kern0pt}length\ {\isasymB}s{\isacharbrackright}{\kern0pt}{\isachardoublequoteclose}\ {\isachardoublequoteopen}{\isasymB}s{\isacharasterisk}{\kern0pt}{\isachardoublequoteclose}\medskip

Intuitively, it is clear that switching the role of blocks and points in our original design to define the dual, results in the dual's incidence matrix being the transpose of the original.

\medskip\isacommand{lemma}\isamarkupfalse%
\ dual{\isacharunderscore}{\kern0pt}incidence{\isacharunderscore}{\kern0pt}mat{\isacharunderscore}{\kern0pt}eq{\isacharunderscore}{\kern0pt}trans{\isacharcolon}{\kern0pt}\ {\isachardoublequoteopen}ordered{\isacharunderscore}{\kern0pt}dual{\isacharunderscore}{\kern0pt}sys{\isachardot}{\kern0pt}N\ {\isacharequal}{\kern0pt}\ N\isactrlsup T{\isachardoublequoteclose}\medskip

The dual system and original system have many symmetries when it comes to properties. For example, if a design has a constant intersect number, its dual is balanced. Similarly, the replication number swaps with block size. This is the first example of a case where it becomes significantly simpler to do this reasoning using the incidence matrix version of the property, demonstrating the power of a linear algebraic representation to formalise even simple lemmas. While a traditional counting proof took 26 lines, the proof could be completed in 10 lines using the incidence matrix. One important resulting statement is that the dual of a constant intersect design is a PBD, located within the \textit{ordered-const-intersect-design} locale.

\medskip\isacommand{lemma}\isamarkupfalse%
\ dual{\isacharunderscore}{\kern0pt}is{\isacharunderscore}{\kern0pt}pbd{\isacharcolon}{\kern0pt}\ \isanewline
\ \ \isakeyword{assumes}\ {\isachardoublequoteopen}{\isacharparenleft}{\kern0pt}{\isasymAnd}\ x\ {\isachardot}{\kern0pt}\ x\ {\isasymin}\ {\isasymV}\ {\isasymLongrightarrow}\ {\isasymB}\ rep\ x\ {\isachargreater}{\kern0pt}\ {\isadigit{0}}{\isacharparenright}{\kern0pt}{\isachardoublequoteclose}
\ \ \isakeyword{and}\ {\isachardoublequoteopen}{\isasymb}\ {\isasymge}\ {\isadigit{2}}{\isachardoublequoteclose}\isanewline
\ \ \isakeyword{shows}\ {\isachardoublequoteopen}pairwise{\isacharunderscore}{\kern0pt}balance\ {\isacharbraceleft}{\kern0pt}{\isadigit{0}}{\isachardot}{\kern0pt}{\isachardot}{\kern0pt}{\isacharless}{\kern0pt}{\isacharparenleft}{\kern0pt}length\ {\isasymB}s{\isacharparenright}{\kern0pt}{\isacharbraceright}{\kern0pt}\ {\isacharparenleft}{\kern0pt}dual{\isacharunderscore}{\kern0pt}blocks\ {\isasymV}\ {\isasymB}s{\isacharparenright}{\kern0pt}\ {\isasymm}{\isachardoublequoteclose}\medskip

\section{The Rank Argument}
\label{sec:rankarg}

\subsection{Mathematical Technique}
The rank argument uses the rank of a matrix to reason about a collection of vectors which form its columns or rows. Given a matrix $A$, its rank $rk(A)$ is defined as the maximum number of linearly independent column (or row) vectors in the matrix. 

Godsil \cite{godsilToolsLinearAlgebra} and Bukh \cite{bukhAlgebraicMethodsCombinatoricsa} both describe the rank argument as one of the foundational techniques in the application of linear algebra to combinatorics. While it can be applied in a number of ways, we present the methodology we use below, given two matrices $N$ and $M$:
\begin{enumerate}
    \item Establish a square matrix of the form $NM$. 
    \item Prove the columns or rows of the matrix are linearly independent, in this case by proving the determinant is non-zero.
    \item Infer that the rank of the matrix must be equal to the number of rows in the matrix.
    \item Using well-known theorems $rk(NM) \leq min(rk(N), rk(M))$, and that for any arbitrary matrix $A$, $rk(A) \leq \text{dim-col}(A)$ and $rk(A) \leq \text{dim-col}(A)$, infer an inequality between the number of rows and columns in our original matrix $N$.
\end{enumerate}

\subsection{Extending Row/Column Operations in Isabelle}

While there are many methods for proving linear independence, in the context of a square matrix showing the determinant is non-zero is a typical textbook approach. When formalising the uniform Fisher's inequality, significant work was required to reason on the determinant formally, where on paper it was a one line description of elementary row and column operations to convert the matrix to an upper triangular form. For this reason, we made several general extensions to the existing row and column operations defined as part of the Gauss-Jordan algorithm formalisation in the AFP \cite{thiemannFormalizingJordanNormal2016}.

On paper, mathematics involving basic application of these elementary operations will often use terminology such as "add row 1 to all other rows in the matrix". To this effect, we extend the operations to enable addition or subtraction of multiple rows and columns at a time, such as the \textit{add-multiple-rows} function. 

\medskip\isacommand{fun}\isamarkupfalse%
\ add{\isacharunderscore}{\kern0pt}multiple{\isacharunderscore}{\kern0pt}rows\ {\isacharcolon}{\kern0pt}{\isacharcolon}{\kern0pt}\ {\isachardoublequoteopen}{\isacharprime}{\kern0pt}a\ {\isacharcolon}{\kern0pt}{\isacharcolon}{\kern0pt}\ semiring{\isacharunderscore}{\kern0pt}{\isadigit{1}}\ {\isasymRightarrow}\ nat\ {\isasymRightarrow}\ nat\ list\ {\isasymRightarrow}\ {\isacharprime}{\kern0pt}a\ mat\ {\isasymRightarrow}\ {\isacharprime}{\kern0pt}a\ mat{\isachardoublequoteclose}\ \isakeyword{where}\isanewline
{\isachardoublequoteopen}add{\isacharunderscore}{\kern0pt}multiple{\isacharunderscore}{\kern0pt}rows\ a\ k\ {\isacharbrackleft}{\kern0pt}{\isacharbrackright}{\kern0pt}\ A\ {\isacharequal}{\kern0pt}\ A{\isachardoublequoteclose}\ {\isacharbar}{\kern0pt}\ \isanewline
{\isachardoublequoteopen}add{\isacharunderscore}{\kern0pt}multiple{\isacharunderscore}{\kern0pt}rows\ a\ k\ {\isacharparenleft}{\kern0pt}l\ {\isacharhash}{\kern0pt}\ ls{\isacharparenright}{\kern0pt}\ A\ {\isacharequal}{\kern0pt}\ {\isacharparenleft}{\kern0pt}addrow\ a\ k\ l\ {\isacharparenleft}{\kern0pt}add{\isacharunderscore}{\kern0pt}multiple{\isacharunderscore}{\kern0pt}rows\ a\ k\ ls\ A{\isacharparenright}{\kern0pt}{\isacharparenright}{\kern0pt}{\isachardoublequoteclose}\medskip

By using lists, we can easily apply induction in addition to pre-existing lemmas on the base operations to formally reason on the functions. In particular, we prove that the determinant remains unchanged from applying these operations. 

\subsection{Rank Argument Formalisation}

To formalise the rank argument, we first prove a simple lemma stating that the rank of the product of two matrices $AB$ must be less than or equal to the minimum of their individual ranks. The proof is straightforward building on results from the rank theory in \cite{thiemannFormalizingJordanNormal2016}. 

\medskip\isacommand{lemma}\isamarkupfalse%
\ rank{\isacharunderscore}{\kern0pt}mat{\isacharunderscore}{\kern0pt}mult{\isacharunderscore}{\kern0pt}lt{\isacharunderscore}{\kern0pt}min{\isacharunderscore}{\kern0pt}rank{\isacharunderscore}{\kern0pt}factor{\isacharcolon}{\kern0pt}\ \isanewline
\ \ \isakeyword{fixes}\ A\ {\isacharcolon}{\kern0pt}{\isacharcolon}{\kern0pt}\ {\isachardoublequoteopen}{\isacharprime}{\kern0pt}a{\isacharcolon}{\kern0pt}{\isacharcolon}{\kern0pt}{\isacharbraceleft}{\kern0pt}conjugatable{\isacharunderscore}{\kern0pt}ordered{\isacharunderscore}{\kern0pt}field{\isacharbraceright}{\kern0pt}\ mat{\isachardoublequoteclose}\isanewline
\ \ \isakeyword{assumes}\ {\isachardoublequoteopen}A\ {\isasymin}\ carrier{\isacharunderscore}{\kern0pt}mat\ n\ m{\isachardoublequoteclose}
\ \ \isakeyword{and}\ {\isachardoublequoteopen}B\ {\isasymin}\ carrier{\isacharunderscore}{\kern0pt}mat\ m\ nc{\isachardoublequoteclose}\ \isanewline
\ \ \isakeyword{shows}\ {\isachardoublequoteopen}vec{\isacharunderscore}{\kern0pt}space{\isachardot}{\kern0pt}rank\ n\ {\isacharparenleft}{\kern0pt}A\ {\isacharasterisk}{\kern0pt}\ B{\isacharparenright}{\kern0pt}\ {\isasymle}\ min\ {\isacharparenleft}{\kern0pt}vec{\isacharunderscore}{\kern0pt}space{\isachardot}{\kern0pt}rank\ n\ A{\isacharparenright}{\kern0pt}\ {\isacharparenleft}{\kern0pt}vec{\isacharunderscore}{\kern0pt}space{\isachardot}{\kern0pt}rank\ m\ B{\isacharparenright}{\kern0pt}{\isachardoublequoteclose}\medskip

Using this lemma and other results on rank, we now prove two versions of the rank argument lemma using the product of a matrix and its transpose. We give below the determinant version of the lemma. 

\medskip\isacommand{lemma}\isamarkupfalse%
\ rank{\isacharunderscore}{\kern0pt}argument{\isacharunderscore}{\kern0pt}det{\isacharcolon}{\kern0pt}\ \isanewline
\ \ \isakeyword{fixes}\ M\ {\isacharcolon}{\kern0pt}{\isacharcolon}{\kern0pt}\ {\isachardoublequoteopen}{\isacharparenleft}{\kern0pt}{\isacharprime}{\kern0pt}c\ {\isacharcolon}{\kern0pt}{\isacharcolon}{\kern0pt}\ {\isacharbraceleft}{\kern0pt}conjugatable{\isacharunderscore}{\kern0pt}ordered{\isacharunderscore}{\kern0pt}field{\isacharbraceright}{\kern0pt}{\isacharparenright}{\kern0pt}\ mat{\isachardoublequoteclose}\isanewline
\ \ \isakeyword{assumes}\ {\isachardoublequoteopen}M\ {\isasymin}\ carrier{\isacharunderscore}{\kern0pt}mat\ x\ y{\isachardoublequoteclose}
\ \ \isakeyword{and}\ {\isachardoublequoteopen}det\ {\isacharparenleft}{\kern0pt}M{\isacharasterisk}{\kern0pt}\ M\isactrlsup T{\isacharparenright}{\kern0pt}\ {\isasymnoteq}\ {\isadigit{0}}{\isachardoublequoteclose}\isanewline
\ \ \isakeyword{shows}\ {\isachardoublequoteopen}x\ {\isasymle}\ y{\isachardoublequoteclose}\medskip

These lemmas can be easily applied to proofs as introduction rules. Effectively, when applied to a proof this version \emph{hides} the rank part of the argument, thus removing the need for someone unfamiliar with the linear algebra libraries in Isabelle to interact with them. In particular, there is no need to interpret a vector space locale, or work within one. This significantly simplifies the proof process. The other version of the lemma gives a more general assumption that is not related to the determinant, enabling greater flexibility if another method is used to prove the rank of the matrix is equal to its column dimension. 

\section{Linear Algebra Method}
\label{sec:linearbound}

\subsection{The Methodology}

The linear algebra method uses similar ideas to the rank argument, however is both more general and as such more widely applicable. In particular it utilises results on dimension and linear independence of vector spaces. 

In general, this method is used to provide an upper bound on the size of a set, what is known as an \textit{extremal combinatorics} problem. There are many fundamental results which use this method in combinatorics, including proofs on graphs, set systems, polynomials, and matrix representations, for which Jukna presents a good overview \cite{juknaExtremalCombinatorics2011}. The general framework for applying it is to associate each object in a set with elements in a vector space of relatively low dimension and show these elements are linearly independent. The fundamental linear algebra bound theorem can then be used to establish an inequality \cite{juknaExtremalCombinatorics2011}. 

\begin{theorem}[Linear Algebra Bound]
    Given a set $v_1, \dots, v_k$ of $k$ linearly independent vectors in a vector space of dimension $m$, then $k \leq m$.
\end{theorem}

Gowers discusses the impact of these dimension tricks in a blog post \cite{gowersDimensionArgumentsCombinatorics2008}. He outlines that for many mathematicians the trick with applying the linear algebra method is often choosing the vector space to use, and establishing the characteristic function which maps our objects to elements in that vector space. However, from a formalisation perspective, the proof process itself can be challenging. We aim to simplify this process, to mirror the mathematical problem, where the challenge remains in knowing when to apply the method while the formal proof process itself is straightforward. The argument itself is formalised in such a way that it is easy to apply to vector representations, such as incidence vectors. We envision that it would be simple to extend or use similar patterns for more complex problems. For example, a common extension out of scope of this formalisation is associating a set to a polynomial, and showing linear independence in the corresponding function space. While this requires a slightly more complicated setup, the resulting proof still uses the same proof pattern as what we formalise here. 

\subsection{Formalisation}

We formalise several versions of the linear algebra method, to allow reasoning directly both on matrix and vector set representations. Firstly, the linear algebra bound presented in theorem (5), has previously been formalised as the lemma \textit{li-le-dim}, in the original vector space locale. Our methods aim to set up a framework to use this fact outside of the context of a vector space. The most general version of the lemma is on a set of vectors:

\medskip\isacommand{lemma}\isamarkupfalse%
\ lin{\isacharunderscore}{\kern0pt}bound{\isacharunderscore}{\kern0pt}arg{\isacharunderscore}{\kern0pt}general{\isacharunderscore}{\kern0pt}set{\isacharcolon}{\kern0pt}\ \isanewline
\ \ \isakeyword{fixes}\ A\ {\isacharcolon}{\kern0pt}{\isacharcolon}{\kern0pt}{\isachardoublequoteopen}{\isacharparenleft}{\kern0pt}{\isacharprime}{\kern0pt}a\ {\isacharcolon}{\kern0pt}{\isacharcolon}{\kern0pt}\ {\isacharbraceleft}{\kern0pt}field{\isacharbraceright}{\kern0pt}{\isacharparenright}{\kern0pt}vec\ set{\isachardoublequoteclose}\isanewline
\ \ \isakeyword{assumes}\ {\isachardoublequoteopen}A\ {\isasymsubseteq}\ carrier{\isacharunderscore}{\kern0pt}vec\ nr{\isachardoublequoteclose}
\ \ \isakeyword{and}\ {\isachardoublequoteopen}vec{\isacharunderscore}{\kern0pt}space{\isachardot}{\kern0pt}lin{\isacharunderscore}{\kern0pt}indpt{\isacharunderscore}{\kern0pt}vs\ nr\ A{\isachardoublequoteclose}\isanewline
\ \ \isakeyword{shows}\ {\isachardoublequoteopen}card\ A\ {\isasymle}\ nr{\isachardoublequoteclose}\medskip

It is clear when applying this method, that the majority of formal proof work is required in the linear independence part of the proof, mirroring the textbook proofs. The most foundational method for proving linear independence is to take an arbitrary linear combination over the vector set and prove that all coefficients must equal 0. Using the vector space definition of linear independence for this methodology involved several layers of unfolding definitions related to linear combinations in calculations. The last version of our linear bound method removes the linear independence criterion of the earlier representation, and replaces it with the matrix version of the linear combination condition, which we prove is equivalent in the vector space context. 

\medskip\isacommand{lemma}\isamarkupfalse%
\ lin{\isacharunderscore}{\kern0pt}bound{\isacharunderscore}{\kern0pt}argument{\isadigit{2}}{\isacharcolon}{\kern0pt}\ \isanewline
\ \ \isakeyword{fixes}\ A\ {\isacharcolon}{\kern0pt}{\isacharcolon}{\kern0pt}\ {\isachardoublequoteopen}{\isacharparenleft}{\kern0pt}{\isacharprime}{\kern0pt}a\ {\isacharcolon}{\kern0pt}{\isacharcolon}{\kern0pt}\ {\isacharbraceleft}{\kern0pt}field{\isacharbraceright}{\kern0pt}{\isacharparenright}{\kern0pt}\ mat{\isachardoublequoteclose}\isanewline
\ \ \isakeyword{assumes}\ {\isachardoublequoteopen}distinct\ {\isacharparenleft}{\kern0pt}cols\ A{\isacharparenright}{\kern0pt}{\isachardoublequoteclose}
\ \ \isakeyword{and}\ {\isachardoublequoteopen}A\ {\isasymin}\ carrier{\isacharunderscore}{\kern0pt}mat\ nr\ nc{\isachardoublequoteclose}\isanewline
\ \ \isakeyword{assumes}\ {\isachardoublequoteopen}{\isasymAnd}\ f{\isachardot}{\kern0pt}\ vec\ nr\ {\isacharparenleft}{\kern0pt}{\isasymlambda}i{\isachardot}{\kern0pt}\ {\isasymSum}\ j\ {\isasymin}\ {\isacharbraceleft}{\kern0pt}{\isadigit{0}}{\isachardot}{\kern0pt}{\isachardot}{\kern0pt}{\isacharless}{\kern0pt}nc{\isacharbraceright}{\kern0pt}\ {\isachardot}{\kern0pt}\ f\ {\isacharparenleft}{\kern0pt}col\ A\ j{\isacharparenright}{\kern0pt}\ {\isacharasterisk}{\kern0pt}\ {\isacharparenleft}{\kern0pt}col\ A\ j{\isacharparenright}{\kern0pt}\ {\isachardollar}{\kern0pt}\ i{\isacharparenright}{\kern0pt}\ {\isacharequal}{\kern0pt}\ {\isadigit{0}}\isactrlsub v\ nr\ {\isasymLongrightarrow}\ \isanewline {\isasymforall}v{\isasymin}\ {\isacharparenleft}{\kern0pt}set\ {\isacharparenleft}{\kern0pt}cols\ A{\isacharparenright}{\kern0pt}{\isacharparenright}{\kern0pt}{\isachardot}{\kern0pt}\ f\ v\ {\isacharequal}{\kern0pt}\ {\isadigit{0}}{\isachardoublequoteclose}\isanewline
\ \ \isakeyword{shows}\ {\isachardoublequoteopen}nc\ {\isasymle}\ nr{\isachardoublequoteclose}\medskip

This removed a significant amount of repeated work from formalisations when applied. As with the rank argument, it additionally simplified formal reasoning by removing the need to understand numerous definitions from the older vector space library and their translation. However, it is worth noting the restriction on the type of the matrix to a field, which is required to use results from the vector space locale. Utilising this lemma still requires the user to implicitly choose the vector space they work in by determining the type of the matrix they are using, and may require some initial setup.

\section{Fisher's Inequality}
\label{sec:fishers}
This section outlines how the above techniques can be applied to prove a number of variations of Fisher's inequality, a consequential result on the relationship between the number of points and blocks in a set system. When discovered, it had immediate implications for applications such as experimental design \cite{babaiLINEARALGEBRAMETHODS1988}, and as a \textit{necessary} condition for design existence, continues to be a valuable tool in both practical and mathematical applications of incidence set systems more generally. Research on generalising the statement is widely credited for the development of linear algebraic methods in combinatorics. The result also is directly used in further proofs, such as a problem Erd\H os presented on three point collinearity \cite{kroesLinearAlgebraMethods}.

\subsection{Incidence Matrix Types}

In order to use both the linear algebra method and rank argument we must be able to reason on a matrix whose elements are of any \textit{field} type. In the context of an \textit{ordered-incidence-system}, an incidence matrix is defined to have integer elements. However, it should be possible to easily lift the incidence matrix to any type which distinguishes zero and one, while preserving basic properties which do not manipulate the incidence matrix elements. For example, the matrix block size property simply counts the number of one's in a column. To formalise Fisher's inequality and its variations, we established a general methodology for 0-1 matrices which could be used to do this transfer between types.

Note firstly, that it is easy to lift integers to a real type using the \textit{of-int} mapping, an injective homomorphism. Building on homomorphism lemmas in the JNF matrix library \cite{thiemannFormalizingJordanNormal2016}, it was straightforward to show certain properties would be preserved. However, the ideal situation is to do this for any such field without repeated work. For example, initial attempts at reasoning over $(\mathbb{Z}/2\mathbb{Z})$ required significant additional set up work, resulting in a Isabelle theory for modulo operations on vectors and integers. This is available as part of the formalisation for completeness, but ultimately was only minimally used in the formalisation of Fisher's inequality. 

Instead, a \textit{of-zero-neq-one} definition is established, providing a simple mapping from any \textit{zero-neq-one} type to another. Under this definition, zero is mapped to zero, and anything else is mapped to one. In a situation where the elements are restricted to just zero and one, such as incidence matrices, the mapping is clearly an injection on that element set. For simplicity, the \textit{lift-01-mat} definition applies this mapping to a matrix.

\medskip\isacommand{definition}\isamarkupfalse%
\ lift{\isacharunderscore}{\kern0pt}{\isadigit{0}}{\isadigit{1}}{\isacharunderscore}{\kern0pt}mat\ {\isacharcolon}{\kern0pt}{\isacharcolon}{\kern0pt}\ {\isachardoublequoteopen}{\isacharprime}{\kern0pt}b\ {\isacharcolon}{\kern0pt}{\isacharcolon}{\kern0pt}\ {\isacharbraceleft}{\kern0pt}zero{\isacharunderscore}{\kern0pt}neq{\isacharunderscore}{\kern0pt}one{\isacharbraceright}{\kern0pt}\ mat\ {\isasymRightarrow}\ {\isacharprime}{\kern0pt}c\ {\isacharcolon}{\kern0pt}{\isacharcolon}{\kern0pt}\ {\isacharbraceleft}{\kern0pt}zero{\isacharunderscore}{\kern0pt}neq{\isacharunderscore}{\kern0pt}one{\isacharbraceright}{\kern0pt}\ mat{\isachardoublequoteclose}\ \isakeyword{where}\ \isanewline
{\isachardoublequoteopen}lift{\isacharunderscore}{\kern0pt}{\isadigit{0}}{\isadigit{1}}{\isacharunderscore}{\kern0pt}mat\ M\ {\isasymequiv}\ map{\isacharunderscore}{\kern0pt}mat\ of{\isacharunderscore}{\kern0pt}zero{\isacharunderscore}{\kern0pt}neq{\isacharunderscore}{\kern0pt}one\ M{\isachardoublequoteclose}\medskip

A number of lemmas are formalised to show that key incidence matrix properties are preserved under the application of the above definition in the \textit{zero-one-matrix} context, similar to existing lemmas on injective homomorphism mappings on matrices such as \textit{of-int}. Note that the result of this definition is also clearly still a \textit{zero-one-matrix} itself, and as such immediately inherits all the properties of this locale. This definition is used in both applications of the linear algebra bound method later in this section.

\subsection{Uniform Fisher's Inequality}

Fisher's inequality was initially proved in 1940 by Fisher \cite{fisherExaminationDifferentPossible1940a} as a result on BIBDs, a very structured incidence system. We give this basic statement in theorem 6.

\begin{theorem}[Uniform Fisher's Inequality]
    Given a $(v, k, \lambda$)-BIBD, with $v$ points, $b$ blocks, uniform block size $k$ and a pairwise points index $\lambda$, we have $v \leq b$. 
\end{theorem}

There are many variations of the proof of Fisher's inequality, but one of the more basic techniques is using the rank argument discussed in section (4). In Isabelle, the theorem is located within the \textit{ordered-bibd} locale, which has the assumptions in its context. 

Notably, the majority of work revolves around calculating the determinant from the diagonal. While the general row and column operations from section (4.2) are straightforward to apply, it remains fiddly to calculate the values of the resulting elements of the matrix after several operations. We reuse results on the properties of the square matrix $NN^T$ for an incidence matrix $N$ of a BIBD, discussed in section (3.5).

The final theorem statement in Isabelle is given below, as well as a sketch of the proof demonstrating how easily the general rank argument lemma can be applied.

\medskip\isacommand{theorem}\isamarkupfalse%
\ Fishers{\isacharunderscore}{\kern0pt}Inequality{\isacharunderscore}{\kern0pt}BIBD{\isacharcolon}{\kern0pt}\ {\isachardoublequoteopen}{\isasymv}\ {\isasymle}\ {\isasymb}{\isachardoublequoteclose}\isanewline
\isacommand{proof}\isamarkupfalse%
\ {\isacharparenleft}{\kern0pt}intro\ rank{\isacharunderscore}{\kern0pt}argument{\isacharunderscore}{\kern0pt}det{\isacharbrackleft}{\kern0pt}of\ {\isachardoublequoteopen}map{\isacharunderscore}{\kern0pt}mat\ real{\isacharunderscore}{\kern0pt}of{\isacharunderscore}{\kern0pt}int\ N{\isachardoublequoteclose}\ {\isachardoublequoteopen}{\isasymv}{\isachardoublequoteclose}\ {\isachardoublequoteopen}{\isasymb}{\isachardoublequoteclose}{\isacharbrackright}{\kern0pt}{\isacharcomma}{\kern0pt}\ simp{\isacharunderscore}{\kern0pt}all{\isacharparenright}{\kern0pt}\isanewline
\ \ \isacommand{show}\isamarkupfalse%
\ {\isachardoublequoteopen}N\ {\isasymin}\ carrier{\isacharunderscore}{\kern0pt}mat\ {\isasymv}\ {\isacharparenleft}{\kern0pt}length\ {\isasymB}s{\isacharparenright}{\kern0pt}{\isachardoublequoteclose}\ \isacommand{using}\isamarkupfalse%
\ blocks{\isacharunderscore}{\kern0pt}list{\isacharunderscore}{\kern0pt}length\ N{\isacharunderscore}{\kern0pt}carrier{\isacharunderscore}{\kern0pt}mat\ \isacommand{by}\isamarkupfalse%
\ simp\isanewline
\ \ \isacommand{let}\isamarkupfalse%
\ {\isacharquery}{\kern0pt}B\ {\isacharequal}{\kern0pt}\ {\isachardoublequoteopen}map{\isacharunderscore}{\kern0pt}mat\ {\isacharparenleft}{\kern0pt}real{\isacharunderscore}{\kern0pt}of{\isacharunderscore}{\kern0pt}int{\isacharparenright}{\kern0pt}\ {\isacharparenleft}{\kern0pt}N\ {\isacharasterisk}{\kern0pt}\ N\isactrlsup T{\isacharparenright}{\kern0pt}{\isachardoublequoteclose}\isanewline
\ \ \isacommand{have}\isamarkupfalse%
\ b{\isacharunderscore}{\kern0pt}split{\isacharcolon}{\kern0pt}\ {\isachardoublequoteopen}{\isacharquery}{\kern0pt}B\ {\isacharequal}{\kern0pt}\ map{\isacharunderscore}{\kern0pt}mat\ {\isacharparenleft}{\kern0pt}real{\isacharunderscore}{\kern0pt}of{\isacharunderscore}{\kern0pt}int{\isacharparenright}{\kern0pt}\ N\ {\isacharasterisk}{\kern0pt}\ {\isacharparenleft}{\kern0pt}map{\isacharunderscore}{\kern0pt}mat\ {\isacharparenleft}{\kern0pt}real{\isacharunderscore}{\kern0pt}of{\isacharunderscore}{\kern0pt}int{\isacharparenright}{\kern0pt}\ N{\isacharparenright}{\kern0pt}\isactrlsup T{\isachardoublequoteclose} \isasymlangle \isatext{proof}\isasymrangle \isanewline
\ \ \isacommand{have}\isamarkupfalse%
\ db{\isacharcolon}{\kern0pt}\ {\isachardoublequoteopen}det\ {\isacharquery}{\kern0pt}B\ {\isacharequal}{\kern0pt}\ {\isacharparenleft}{\kern0pt}{\isasymr}\ {\isacharplus}{\kern0pt}\ {\isasymLambda}\ {\isacharasterisk}{\kern0pt}\ {\isacharparenleft}{\kern0pt}{\isasymv}\ {\isacharminus}{\kern0pt}\ {\isadigit{1}}{\isacharparenright}{\kern0pt}{\isacharparenright}{\kern0pt}{\isacharasterisk}{\kern0pt}\ {\isacharparenleft}{\kern0pt}{\isasymr}\ {\isacharminus}{\kern0pt}\ {\isasymLambda}{\isacharparenright}{\kern0pt}{\isacharcircum}{\kern0pt}{\isacharparenleft}{\kern0pt}{\isasymv}\ {\isacharminus}{\kern0pt}\ {\isadigit{1}}{\isacharparenright}{\kern0pt}{\isachardoublequoteclose} \isasymlangle \isatext{proof}\isasymrangle \isanewline
\ \ \isacommand{have}\isamarkupfalse%
\ lhn{\isadigit{0}}{\isacharcolon}{\kern0pt}\ {\isachardoublequoteopen}{\isacharparenleft}{\kern0pt}{\isasymr}\ {\isacharplus}{\kern0pt}\ {\isasymLambda}\ {\isacharasterisk}{\kern0pt}\ {\isacharparenleft}{\kern0pt}{\isasymv}\ {\isacharminus}{\kern0pt}\ {\isadigit{1}}{\isacharparenright}{\kern0pt}{\isacharparenright}{\kern0pt}\ {\isachargreater}{\kern0pt}\ {\isadigit{0}}{\isachardoublequoteclose} \isasymlangle \isatext{proof}\isasymrangle\isanewline
\ \ \isacommand{have}\isamarkupfalse%
\ {\isachardoublequoteopen}{\isacharparenleft}{\kern0pt}{\isasymr}\ {\isacharminus}{\kern0pt}\ {\isasymLambda}{\isacharparenright}{\kern0pt}\ {\isachargreater}{\kern0pt}\ {\isadigit{0}}{\isachardoublequoteclose} \isasymlangle \isatext{proof}\isasymrangle \isanewline
\ \ \isacommand{then}\isamarkupfalse%
\ \isacommand{have}\isamarkupfalse%
\ det{\isacharunderscore}{\kern0pt}not{\isacharunderscore}{\kern0pt}{\isadigit{0}}{\isacharcolon}{\kern0pt}\ \ {\isachardoublequoteopen}det\ {\isacharquery}{\kern0pt}B\ {\isasymnoteq}\ {\isadigit{0}}{\isachardoublequoteclose}\ \isasymlangle \isatext{proof}\isasymrangle \isanewline
\ \ \isacommand{thus}\isamarkupfalse%
\ {\isachardoublequoteopen}det\ {\isacharparenleft}{\kern0pt}of{\isacharunderscore}{\kern0pt}int{\isacharunderscore}{\kern0pt}hom{\isachardot}{\kern0pt}mat{\isacharunderscore}{\kern0pt}hom\ N\ {\isacharasterisk}{\kern0pt}\ {\isacharparenleft}{\kern0pt}of{\isacharunderscore}{\kern0pt}int{\isacharunderscore}{\kern0pt}hom{\isachardot}{\kern0pt}mat{\isacharunderscore}{\kern0pt}hom\ N{\isacharparenright}{\kern0pt}\isactrlsup T{\isacharparenright}{\kern0pt}\ {\isasymnoteq}\ {\isacharparenleft}{\kern0pt}{\isadigit{0}}{\isacharcolon}{\kern0pt}{\isacharcolon}{\kern0pt}\ real{\isacharparenright}{\kern0pt}{\isachardoublequoteclose}\ \isasymlangle \isatext{proof}\isasymrangle \isanewline
\isacommand{qed}\isamarkupfalse%
\medskip

\subsection{Variations: The Odd Town Problem}
The classic introductory problem for the linear algebra method in numerous lecture notes on the subject is that of even and odd towns \cite{babaiLINEARALGEBRAMETHODS1988}. In particular, the odd town problem presents a bound similar to Fisher's inequality. It aims to identify the maximum number of clubs a town can have, given the conditions that the intersection of any two clubs is even, and every club has an odd number of people. We present this formally below. 
\begin{theorem}
    Given a set $\mathcal{V} = \{0..<n\}$, and a collection $\mathcal{B}$ of $m$ subsets of $\mathcal{V}$ of odd size, such that for any $i, j$ where $i \neq j$, $B_i \cap B_j$ is even, $\mathcal{B}$ can be at most of size $n$, i.e. $m \leq n$. 
\end{theorem}

We can easily represent this problem formally by building on our locales for ordered incidence systems, and definitions on block size and intersection number.

\medskip\isacommand{locale}\isamarkupfalse%
\ odd{\isacharunderscore}{\kern0pt}town\ {\isacharequal}{\kern0pt}\ ordered{\isacharunderscore}{\kern0pt}design\ {\isacharplus}{\kern0pt}\ \isanewline
\ \ \isakeyword{assumes}\ odd{\isacharunderscore}{\kern0pt}groups{\isacharcolon}{\kern0pt}\ {\isachardoublequoteopen}bl\ {\isasymin}{\isacharhash}{\kern0pt}\ {\isasymB}\ {\isasymLongrightarrow}\ odd\ {\isacharparenleft}{\kern0pt}card\ bl{\isacharparenright}{\kern0pt}{\isachardoublequoteclose}\isanewline
\ \ \isakeyword{and}\ even{\isacharunderscore}{\kern0pt}inters{\isacharcolon}{\kern0pt}\ {\isachardoublequoteopen}bl{\isadigit{1}}\ {\isasymin}{\isacharhash}{\kern0pt}\ {\isasymB}\ {\isasymLongrightarrow}\ bl{\isadigit{2}}\ {\isasymin}{\isacharhash}{\kern0pt}\ {\isacharparenleft}{\kern0pt}{\isasymB}\ {\isacharminus}{\kern0pt}\ {\isacharbraceleft}{\kern0pt}{\isacharhash}{\kern0pt}bl{\isadigit{1}}{\isacharhash}{\kern0pt}{\isacharbraceright}{\kern0pt}{\isacharparenright}{\kern0pt}\ \ {\isasymLongrightarrow}\ even\ {\isacharparenleft}{\kern0pt}bl{\isadigit{1}}\ {\isacharbar}{\kern0pt}{\isasyminter}{\isacharbar}{\kern0pt}\ bl{\isadigit{2}}{\isacharparenright}{\kern0pt}{\isachardoublequoteclose}\medskip

Additionally, the condition for the intersection between any two blocks of odd size to be even in turn implies that no two blocks can be the same. In other words, this is a simple design, which we can prove through the sublocale command.

In this case, given our incidence vectors are 0-1 vectors, we prove they are linearly independent over the vector space $(\mathbb{Z}/2\mathbb{Z})^n$. This clearly has dimension $n$, our required upper bound. Note that the scalar product of an incidence vector with itself is equal to the block size which is odd: $v_i \cdot v_i = 1\; (\text{mod}\, 2)$. Similarly for any two different vectors, the scalar product is the intersection number, which is even: $i \neq j \longrightarrow v_i \cdot v_j = 0 \; (\text{mod} \,2)$. These results are formalised in separate lemma statements using the matrix definitions of block size and intersection number. The proof of linear independence proceeds by taking an arbitrary linear combination equal to 0, and multiplying it by an incidence vector $v$ from our set. Using the results on the scalar product, we can prove that the coefficient of $v$ must be 0, as required. 

The calculations above are trivial in $(\mathbb{Z}/2\mathbb{Z})$, however this presents the first challenge of formalising the proof. Given Isabelle's foundations in simple type theory, some creativity is required to define a finite field type. For the purpose of this proof, we went with the formalisation presented in the Berlekamp Zassenhaus AFP entry \cite{divasonVerifiedImplementationBerlekamp2020}. This uses the lifting and transfer methodology \cite{huffmanLiftingTransferModular2013} to prove the resulting type \textit{'a mod-ring} is a field, which is required for our linear bound argument lemma. Using the \textit{lift-01-mat} definition from section (6.1), it is possible to easily lift our incidence matrix to this type. 

The formal statement of the lemma is within the odd clubs locale:

\medskip\isacommand{lemma}\isamarkupfalse%
\ upper{\isacharunderscore}{\kern0pt}bound{\isacharunderscore}{\kern0pt}clubs{\isacharcolon}{\kern0pt}\ \isanewline
\ \ \isakeyword{assumes}\ {\isachardoublequoteopen}CARD{\isacharparenleft}{\kern0pt}{\isacharprime}{\kern0pt}b{\isacharcolon}{\kern0pt}{\isacharcolon}{\kern0pt}prime{\isacharunderscore}{\kern0pt}card{\isacharparenright}{\kern0pt}\ {\isacharequal}{\kern0pt}\ {\isadigit{2}}{\isachardoublequoteclose}\isanewline
\ \ \isakeyword{shows}\ {\isachardoublequoteopen}{\isasymb}\ {\isasymle}\ {\isasymv}{\isachardoublequoteclose}\isanewline
\isacommand{proof}\isamarkupfalse%
\ {\isacharminus}{\kern0pt}\isanewline
\ \ \isacommand{have}\isamarkupfalse%
\ cb{\isadigit{2}}{\isacharcolon}{\kern0pt}\ {\isachardoublequoteopen}CARD{\isacharparenleft}{\kern0pt}{\isacharprime}{\kern0pt}b{\isacharparenright}{\kern0pt}\ {\isacharequal}{\kern0pt}\ {\isadigit{2}}{\isachardoublequoteclose}\ \isacommand{using}\isamarkupfalse%
\ assms\ \isacommand{by}\isamarkupfalse%
\ simp\isanewline
\ \ \isacommand{define}\isamarkupfalse%
\ N{\isadigit{2}}\ {\isacharcolon}{\kern0pt}{\isacharcolon}{\kern0pt}\ {\isachardoublequoteopen}{\isacharprime}{\kern0pt}b\ mod{\isacharunderscore}{\kern0pt}ring\ mat{\isachardoublequoteclose}\ \isakeyword{where}\ {\isachardoublequoteopen}N{\isadigit{2}}\ {\isasymequiv}\ lift{\isacharunderscore}{\kern0pt}{\isadigit{0}}{\isadigit{1}}{\isacharunderscore}{\kern0pt}mat\ N{\isachardoublequoteclose}\isanewline
\ \ \isacommand{show}\isamarkupfalse%
\ {\isacharquery}{\kern0pt}thesis\ \isacommand{proof}\isamarkupfalse%
\ {\isacharparenleft}{\kern0pt}intro\ lin{\isacharunderscore}{\kern0pt}bound{\isacharunderscore}{\kern0pt}argument{\isadigit{2}}{\isacharbrackleft}{\kern0pt}of\ {\isachardoublequoteopen}N{\isadigit{2}}{\isachardoublequoteclose}{\isacharbrackright}{\kern0pt}{\isacharparenright}{\kern0pt}\isanewline
\ \ \ \ \isacommand{show}\isamarkupfalse%
\ {\isachardoublequoteopen}distinct\ {\isacharparenleft}{\kern0pt}cols\ {\isacharparenleft}{\kern0pt}N{\isadigit{2}}{\isacharparenright}{\kern0pt}{\isacharparenright}{\kern0pt}{\isachardoublequoteclose}\ \isasymlangle \isatext{proof}\isasymrangle \isanewline
\ \ \ \ \isacommand{show}\isamarkupfalse%
\ n{\isadigit{2}}cm{\isacharcolon}{\kern0pt}{\isachardoublequoteopen}N{\isadigit{2}}\ {\isasymin}\ carrier{\isacharunderscore}{\kern0pt}mat\ {\isasymv}\ {\isasymb}{\isachardoublequoteclose}\ \isasymlangle \isatext{proof}\isasymrangle \isanewline
\ \ \ \ \isacommand{show}\isamarkupfalse%
\ {\isachardoublequoteopen}{\isasymAnd}f{\isachardot}{\kern0pt}\ vec\ {\isasymv}\ {\isacharparenleft}{\kern0pt}{\isasymlambda}i{\isachardot}{\kern0pt}\ {\isasymSum}j\ {\isacharequal}{\kern0pt}\ {\isadigit{0}}{\isachardot}{\kern0pt}{\isachardot}{\kern0pt}{\isacharless}{\kern0pt}{\isasymb}{\isachardot}{\kern0pt}\ f\ {\isacharparenleft}{\kern0pt}col\ N{\isadigit{2}}\ j{\isacharparenright}{\kern0pt}\ {\isacharasterisk}{\kern0pt}\ {\isacharparenleft}{\kern0pt}col\ N{\isadigit{2}}\ j{\isacharparenright}{\kern0pt}\ {\isachardollar}{\kern0pt}\ i{\isacharparenright}{\kern0pt}\ {\isacharequal}{\kern0pt}\ {\isadigit{0}}\isactrlsub v\ {\isasymv}\ {\isasymLongrightarrow}\ \isanewline 
\ \ \ \ \ {\isasymforall}v{\isasymin}set\ {\isacharparenleft}{\kern0pt}cols\ N{\isadigit{2}}{\isacharparenright}{\kern0pt}{\isachardot}{\kern0pt}\ f\ v\ {\isacharequal}{\kern0pt}\ {\isadigit{0}}{\isachardoublequoteclose}\isanewline
\ \ \ \ \isacommand{proof}\isamarkupfalse%
\ {\isacharparenleft}{\kern0pt}auto{\isacharparenright}{\kern0pt}\isanewline
\ \ \ \ \ \ \isacommand{fix}\isamarkupfalse%
\ f\ v\ \ \isacommand{assume}\isamarkupfalse%
\ vin{\isacharcolon}{\kern0pt}\ {\isachardoublequoteopen}v\ {\isasymin}\ set\ {\isacharparenleft}{\kern0pt}cols\ N{\isadigit{2}}{\isacharparenright}{\kern0pt}{\isachardoublequoteclose}\isanewline
\ \ \ \ \ \ \isacommand{assume}\isamarkupfalse%
\ eq{\isadigit{0}}{\isacharcolon}{\kern0pt}\ {\isachardoublequoteopen}vec\ {\isasymv}\ {\isacharparenleft}{\kern0pt}{\isasymlambda}i{\isachardot}{\kern0pt}\ {\isasymSum}j\ {\isacharequal}{\kern0pt}\ {\isadigit{0}}{\isachardot}{\kern0pt}{\isachardot}{\kern0pt}{\isacharless}{\kern0pt}length\ {\isasymB}s{\isachardot}{\kern0pt}\ f\ {\isacharparenleft}{\kern0pt}col\ N{\isadigit{2}}\ j{\isacharparenright}{\kern0pt}\ {\isacharasterisk}{\kern0pt}\ {\isacharparenleft}{\kern0pt}col\ N{\isadigit{2}}\ j{\isacharparenright}{\kern0pt}\ {\isachardollar}{\kern0pt}\ i{\isacharparenright}{\kern0pt}\ {\isacharequal}{\kern0pt}\ {\isadigit{0}}\isactrlsub v\ {\isasymv}{\isachardoublequoteclose}\ \isanewline
\ \ \ \ \ \ \ \isasymlangle \isatext{Isar linear independence proof details omitted}\isasymrangle \isanewline
\ \ \ \ \ \ \isacommand{then}\isamarkupfalse%
\ \isacommand{show}\isamarkupfalse%
\ {\isachardoublequoteopen}f\ v\ {\isacharequal}{\kern0pt}\ {\isadigit{0}}{\isachardoublequoteclose} \isasymlangle \isatext{proof}\isasymrangle \isanewline
\ \ \ \ \isacommand{qed}\isamarkupfalse%
\isanewline
\ \ \isacommand{qed}\isamarkupfalse%
\isanewline
\isacommand{qed}\isamarkupfalse%
\medskip

The proof sketch above clearly shows the structure of the formal proof using the linear bound lemma from section (5.2). The initial part of the formalisation applies the \textit{lift-01-mat} definition to establish \textit{N2}, an incidence matrix of the type \textit{'b mod-ring mat}. We can then apply the linear bound method directly, resulting in three goals. The first two goals are straightforward using lemmas on the \textit{lift-01-mat} definition and known facts on the incidence matrix $N$ given the locale context. The proof of the final goal closely follows the mathematical proof outlined earlier. The details omitted from the sketch above include manipulating a summation and utilising the formal results on the scalar product of two incidence vectors. 

\subsection{Generalised Fisher's Inequality}
The generalised version of Fisher's inequality was first proved by Majumdar \cite{majumdarTheoremsCombinatoricsRelating1953}, after work by Bose, de Bruijn and Erd\H os on Fisher's initial statement. It was Bose \cite{boseNoteFisherInequality1949} who first presented the linear algebraic approach to prove a more general version of the theorem, which Majumdar then built on. Notably, the generalised version of Fisher's inequality is presented on families of intersecting sets, rather than using design theoretic terminology. 

\begin{theorem}
    Let $A_1, ..., A_m$ be distinct subsets of ${1,..., n}$ such that $|A_i \cap A_j| = k$ for some fixed $1 \le k \le n$ and every $i \neq j$. Then $m \le n$.
\end{theorem}

This is a familiar statement in comparison to both earlier variations, clearly putting a bound on any incidence set system with the constant intersect property. It is both more general than the uniform version of Fisher's, instead only constraining the intersect number rather than the points index, and has no condition on the size of blocks as in the odd town statement, while the intersect condition is stronger.

The proof we follow was presented by Jukna \cite{juknaExtremalCombinatorics2011}, building on initial work by Babai and Frankl \cite{babaiLINEARALGEBRAMETHODS1988}. On paper, the proof takes up less than half a page and is dominated by a linear independence proof. It utilises similar results on the scalar product of incidence vectors as in the odd town proof, however as we are no longer working in $(\mathbb{Z}/2\mathbb{Z})$, the calculations require more creativity to prove that each coefficient must be equal to 0 in an arbitrary linear combination equal to the 0 vector. We begin by taking the scalar product of this linear combination with itself (rather than a singular incidence vector as in the odd town proof). Through manipulating the resultant summations, we are able to arrive at an equation where each term must clearly be positive. Using the intersection condition $|A_i| \geq k$ for all $i$ and $|A_i| = k$ for at most one $i$, we are able to conclude that each coefficient must be 0.

Unlike the odd-town proof, we first prove trivial cases for when there are less than two blocks or a zero intersect number. However, we are similarly able to once again use the \textit{lift-01-mat} definition to establish an incidence matrix with real elements \textit{NR}. We proceed with the main part of the proof by applying the same linear bound argument. 

As in the odd town case, the proofs of the first two properties are quick. While it would be possible to simplify these properties out entirely, it is useful to be able to reference these facts directly in the linear independence proof. Previously proven facts on the relationship between the scalar product, block size and intersection numbers can also be reused given lemmas on the preservation of incidence matrix properties under the \textit{lift-01-mat} definition. The linear independence proof itself is more complicated than the odd town proof, having two main components. Firstly, we manipulate a summation into a split form through the initial scalar product step, resulting in the \textit{finally} statement in the proof sketch below. We then reason on why the coefficients must evaluate to 0 using this split form, through the lemma \textit{sum-split-coeffs-0}. Here, we can see the benefits of using the modified linear bound lemma, as summation reasoning for this and the odd town proof has minimal repetition. A sketch of the final proof in Isabelle, located in the \textit{simp-ordered-const-intersect-design} locale, is given below:

\medskip\isacommand{theorem}\isamarkupfalse%
\ general{\isacharunderscore}{\kern0pt}fishers{\isacharunderscore}{\kern0pt}inequality{\isacharcolon}{\kern0pt}\ \ {\isachardoublequoteopen}{\isasymb}\ {\isasymle}\ {\isasymv}{\isachardoublequoteclose}\isanewline
\isacommand{proof}\isamarkupfalse%
\ {\isacharparenleft}{\kern0pt}cases\ {\isachardoublequoteopen}{\isasymm}\ {\isacharequal}{\kern0pt}\ {\isadigit{0}}\ {\isasymor}\ {\isasymb}\ {\isacharequal}{\kern0pt}\ {\isadigit{1}}{\isachardoublequoteclose}{\isacharparenright}{\kern0pt}\isanewline
\ \ \isacommand{case}\isamarkupfalse%
\ True\ \isacommand{then}\isamarkupfalse%
\ \isacommand{show}\isamarkupfalse%
\ {\isacharquery}{\kern0pt}thesis\ \isasymlangle \isatext{proof}\isasymrangle \isanewline
\isacommand{next}\isamarkupfalse%
\isanewline
\ \ \isacommand{case}\isamarkupfalse%
\ False\isanewline
\ \ \isacommand{define}\isamarkupfalse%
\ NR\ {\isacharcolon}{\kern0pt}{\isacharcolon}{\kern0pt}\ {\isachardoublequoteopen}real\ mat{\isachardoublequoteclose}\ \isakeyword{where}\ {\isachardoublequoteopen}NR\ {\isasymequiv}\ lift{\isacharunderscore}{\kern0pt}{\isadigit{0}}{\isadigit{1}}{\isacharunderscore}{\kern0pt}mat\ N{\isachardoublequoteclose}\isanewline
\ \ \isacommand{show}\isamarkupfalse%
\ {\isacharquery}{\kern0pt}thesis\ \isacommand{proof}\isamarkupfalse%
\ {\isacharparenleft}{\kern0pt}intro\ lin{\isacharunderscore}{\kern0pt}bound{\isacharunderscore}{\kern0pt}argument{\isadigit{2}}{\isacharbrackleft}{\kern0pt}of\ NR{\isacharbrackright}{\kern0pt}{\isacharparenright}{\kern0pt}\isanewline
\ \ \ \ \isacommand{show}\isamarkupfalse%
\ {\isachardoublequoteopen}distinct\ {\isacharparenleft}{\kern0pt}cols\ NR{\isacharparenright}{\kern0pt}{\isachardoublequoteclose}\ \isasymlangle \isatext{proof}\isasymrangle \isanewline
\ \ \ \ \isacommand{show}\isamarkupfalse%
\ nrcm{\isacharcolon}{\kern0pt}\ {\isachardoublequoteopen}NR\ {\isasymin}\ carrier{\isacharunderscore}{\kern0pt}mat\ {\isasymv}\ {\isasymb}{\isachardoublequoteclose}\ \isasymlangle \isatext{proof}\isasymrangle \isanewline
\ \ \ \ \isacommand{show}\isamarkupfalse%
\ {\isachardoublequoteopen}{\isasymAnd}f{\isachardot}{\kern0pt}\ vec\ {\isasymv}\ {\isacharparenleft}{\kern0pt}{\isasymlambda}i{\isachardot}{\kern0pt}\ {\isasymSum}j\ {\isacharequal}{\kern0pt}\ {\isadigit{0}}{\isachardot}{\kern0pt}{\isachardot}{\kern0pt}{\isacharless}{\kern0pt}{\isasymb}{\isachardot}{\kern0pt}\ f\ {\isacharparenleft}{\kern0pt}col\ NR\ j{\isacharparenright}{\kern0pt}\ {\isacharasterisk}{\kern0pt}\ {\isacharparenleft}{\kern0pt}col\ NR\ j{\isacharparenright}{\kern0pt}\ {\isachardollar}{\kern0pt}\ i{\isacharparenright}{\kern0pt}\ {\isacharequal}{\kern0pt}\ {\isadigit{0}}\isactrlsub v\ {\isasymv}\ {\isasymLongrightarrow}\ \isanewline \ \ \ \ \ {\isasymforall}v{\isasymin}set\ {\isacharparenleft}{\kern0pt}cols\ NR{\isacharparenright}{\kern0pt}{\isachardot}{\kern0pt}\ f\ v\ {\isacharequal}{\kern0pt}\ {\isadigit{0}}{\isachardoublequoteclose}\isanewline
\ \ \ \ \isacommand{proof}\isamarkupfalse%
\ {\isacharparenleft}{\kern0pt}intro\ ballI{\isacharparenright}{\kern0pt}\isanewline
\ \ \ \ \ \ \isacommand{fix}\isamarkupfalse%
\ f\ v \ \isacommand{assume}\isamarkupfalse%
\ vin{\isacharcolon}{\kern0pt}\ {\isachardoublequoteopen}v\ {\isasymin}\ set\ {\isacharparenleft}{\kern0pt}cols\ NR{\isacharparenright}{\kern0pt}{\isachardoublequoteclose}\isanewline
\ \ \ \ \ \ \isacommand{assume}\isamarkupfalse%
\ eq{\isadigit{0}}{\isacharcolon}{\kern0pt}\ {\isachardoublequoteopen}vec\ {\isasymv}\ {\isacharparenleft}{\kern0pt}{\isasymlambda}i{\isachardot}{\kern0pt}\ {\isasymSum}j\ {\isacharequal}{\kern0pt}\ {\isadigit{0}}{\isachardot}{\kern0pt}{\isachardot}{\kern0pt}{\isacharless}{\kern0pt}{\isasymb}{\isachardot}{\kern0pt}\ f\ {\isacharparenleft}{\kern0pt}col\ NR\ j{\isacharparenright}{\kern0pt}\ {\isacharasterisk}{\kern0pt}\ col\ NR\ j\ {\isachardollar}{\kern0pt}\ i{\isacharparenright}{\kern0pt}\ {\isacharequal}{\kern0pt}\ {\isadigit{0}}\isactrlsub v\ {\isasymv}{\isachardoublequoteclose}\isanewline
\ \ \ \ \ \ \isacommand{define}\isamarkupfalse%
\ c\ \isakeyword{where}\ {\isachardoublequoteopen}c\ {\isasymequiv}\ {\isacharparenleft}{\kern0pt}{\isasymlambda}\ j{\isachardot}{\kern0pt}\ f\ {\isacharparenleft}{\kern0pt}col\ NR\ j{\isacharparenright}{\kern0pt}{\isacharparenright}{\kern0pt}{\isachardoublequoteclose}\isanewline
\ \ \ \ \ \ \isasymlangle \isatext{Isar linear independence proof details omitted}\isasymrangle \isanewline
\ \ \ \ \ \ \isacommand{finally}\isamarkupfalse%
\ \isacommand{have}\isamarkupfalse%
\ sum{\isacharunderscore}{\kern0pt}rep{\isacharcolon}{\kern0pt}\ {\isachardoublequoteopen}{\isadigit{0}}\ {\isacharequal}{\kern0pt}\ {\isacharparenleft}{\kern0pt}{\isasymSum}\ j\ {\isasymin}\ {\isacharbraceleft}{\kern0pt}{\isadigit{0}}{\isachardot}{\kern0pt}{\isachardot}{\kern0pt}{\isacharless}{\kern0pt}{\isasymb}{\isacharbraceright}{\kern0pt}\ {\isachardot}{\kern0pt}\ {\isacharparenleft}{\kern0pt}c\ j{\isacharparenright}{\kern0pt}{\isacharcircum}{\kern0pt}{\isadigit{2}}\ {\isacharasterisk}{\kern0pt}\ {\isacharparenleft}{\kern0pt}{\isacharparenleft}{\kern0pt}card\ {\isacharparenleft}{\kern0pt}{\isasymB}s\ {\isacharbang}{\kern0pt}\ j{\isacharparenright}{\kern0pt}{\isacharparenright}{\kern0pt}{\isacharminus}{\kern0pt}\ {\isacharparenleft}{\kern0pt}int\ {\isasymm}{\isacharparenright}{\kern0pt}{\isacharparenright}{\kern0pt}{\isacharparenright}{\kern0pt}\ {\isacharplus}{\kern0pt}\ \isanewline
\ \ \ \ \ \ \ \ \ {\isasymm}\ {\isacharasterisk}{\kern0pt}\ {\isacharparenleft}{\kern0pt}{\isacharparenleft}{\kern0pt}{\isasymSum}\ j\ {\isasymin}\ {\isacharbraceleft}{\kern0pt}{\isadigit{0}}{\isachardot}{\kern0pt}{\isachardot}{\kern0pt}{\isacharless}{\kern0pt}{\isasymb}{\isacharbraceright}{\kern0pt}\ {\isachardot}{\kern0pt}\ c\ j{\isacharparenright}{\kern0pt}{\isacharcircum}{\kern0pt}{\isadigit{2}}{\isacharparenright}{\kern0pt}{\isachardoublequoteclose}\ \isacommand{by}\isamarkupfalse%
\ {\isacharparenleft}{\kern0pt}simp\ add{\isacharcolon}{\kern0pt}\ algebra{\isacharunderscore}{\kern0pt}simps{\isacharparenright}{\kern0pt}\isanewline
\ \ \ \ \ \ \isacommand{thus}\isamarkupfalse%
\ {\isachardoublequoteopen}f\ v\ {\isacharequal}{\kern0pt}\ {\isadigit{0}}{\isachardoublequoteclose}\ \isacommand{using}\isamarkupfalse%
\ sum{\isacharunderscore}{\kern0pt}split{\isacharunderscore}{\kern0pt}coeffs{\isacharunderscore}{\kern0pt}{\isadigit{0}}{\isacharbrackleft}{\kern0pt}of\ {\isachardoublequoteopen}j{\isacharprime}{\kern0pt}{\isachardoublequoteclose}\ {\isachardoublequoteopen}c{\isachardoublequoteclose}{\isacharbrackright}{\kern0pt}\ \isasymlangle \isatext{proof}\isasymrangle \isanewline
\ \ \ \ \isacommand{qed}\isamarkupfalse%
\isanewline
\ \ \isacommand{qed}\isamarkupfalse%
\isanewline
\isacommand{qed}\isamarkupfalse%
\medskip

\subsection{Dual of Generalised Fisher's}

A brief analysis of the uniform version of Fisher's demonstrates that it is in fact a restricted version of the dual of the generalised version, with the inequality flipped. In design theory, it is significantly more common to reason about PBDs than it is to reason about designs with constant intersect numbers. Using the dual design formalisation outlined in section (3.6), we've previously proven that the dual of a PBD with $v$ points and $b$ blocks is a constant intersect design with $b$ points and $v$ blocks. With the added condition of incompleteness, the dual is a simple constant intersect design, which is the context we proved the generalised version of Fisher's inequality in. Using locale interpretation, it is simple to apply Fisher's inequality to get the dual bound within the pairwise balanced locale, as desired:

\medskip\isacommand{corollary}\isamarkupfalse%
\ general{\isacharunderscore}{\kern0pt}nonuniform{\isacharunderscore}{\kern0pt}fishers{\isacharcolon}{\kern0pt}\ \isanewline
\ \ \isakeyword{assumes}\ {\isachardoublequoteopen}{\isasymLambda}\ {\isachargreater}{\kern0pt}\ {\isadigit{0}}{\isachardoublequoteclose}\ \isakeyword{and}\ {\isachardoublequoteopen}{\isasymAnd}\ bl{\isachardot}{\kern0pt}\ bl\ {\isasymin}{\isacharhash}{\kern0pt}\ {\isasymB}\ {\isasymLongrightarrow}\ incomplete{\isacharunderscore}{\kern0pt}block\ bl{\isachardoublequoteclose}\ \isanewline
\ \ \isakeyword{shows}\ {\isachardoublequoteopen}{\isasymv}\ {\isasymle}\ {\isasymb}{\isachardoublequoteclose}\medskip

Note that this technically implies the uniform version of Fisher's on BIBDs which we started with. However, the rank argument remains a valuable tool, and hence we include both versions of the proof in our final formalisation. 

\section{Discussion}

There are several valuable discussion points from this formalisation, of which we focus on two areas. Firstly, the challenges in formalising proofs utilising techniques from multiple mathematical fields and past formalisations, and secondly a comparison of the linear algebraic and traditional combinatorial proof approaches in a formal environment.

\subsection{Integrating Multiple Formal Libraries}

Modern mathematics routinely deals with proofs which draw on unexpected techniques from other mathematical fields. This is particularly true of combinatorics, as discussed in section (1), where a barrier to formalisation in the past has been the need for established libraries in other fields of mathematics first. As formal libraries expand, it is anticipated it will routinely become more common to draw on formal results from a variety of fields and past formalisations. This formalisation was at the intersection of just two fields, using straightforward results from linear algebra as applied to combinatorics. Yet it highlights some of the barriers that still exist, particularly in older well established proof assistants such as Isabelle which are both benefited and hindered by decades of contributions. We summarise the challenges as follows: 
\begin{romanenumerate}
    \item Managing multiple representations of the same concept in established libraries, and the transfer of results between these representations. 
    \item The flexibility of these representations for extensions and future applications. 
    \item Finding the necessary results as needed, including often routine lemmas.
\end{romanenumerate}

Point (i) is likely the most relevant factor from a mathematician's perspective. In this formalisation matrices, vectors, vector spaces, and prime fields are notable examples. For a less experienced user, one would imagine that even the initial choice between representations may prove a barrier to proving results utilising these theories. In Isabelle, a survey of the AFP shows the JNF library \cite{thiemannFormalizingJordanNormal2016} is quickly becoming the matrix library of choice for future formalisations, and further documentation summarising the current usage of certain libraries would likely prove useful. The transfer of results between representations remains crucial, both for existing libraries, and in the case of the creation of a new, more effective representations for a structure. The lifting and transfer libraries \cite{huffmanLiftingTransferModular2013} are a powerful tool here, and are well used in the matrix and vector libraries, however can be fiddly to set up. This formalisation experimented with using transfer rules to move between integer matrices under modular arithmetic and matrices of type \textit{'a mod-ring}, a type originally defined using lift definitions \cite{divasonVerifiedImplementationBerlekamp2020}. Ultimately, a different approach was used and it was simpler to just prove the few equivalence results required directly. 

The flexibility of past formalisations is essential to reducing duplication. In Isabelle, locales have proven to be valuable towards these efforts. This was again demonstrated in this formalisation, by the ease at which our existing incidence system locales from past work \cite{edmondsModularFirstFormalisation2021} could be extended and adapted. Similarly, they enable the transfer of results between vector space definitions. However, one notable limitation of locales in this formalisation was the use of definitions outside of the locale context. Inherited definitions could not be accessed directly, requiring an interpretation of the locale, while others could be such as \textit{vec-space.rank}. To overcome this in our methodology, we defined a trivial equivalent definition, which ideally would not be needed.

Lastly (iii) addresses one of the most significant challenges of this formalisation, which was both determining the existence of and sourcing, sometimes trivial, results. In particular, for linear algebra in Isabelle, results are scattered across many AFP entries. Numerous entries which build on the JNF Matrix representation include a theory with a significant number of basic but useful results on vectors and matrices, but these can be difficult to find. This also means AFP entries have a growing number of dependencies and this formalisation will likely add yet another entry to the convoluted network of linear algebra results. Search tools aiming to answer this research question such as SErAPIS proved extremely useful in navigating these libraries when beginning the formalisation project. 

\subsection{Formalising the Linear Algebraic Method }
The linear algebraic method is ultimately a very simple concept, but one that can be quite difficult to apply. Here we discuss how working with linear algebra in a formal environment compares to a traditional combinatorial approach and the joint focus on both proof techniques and theorems.

Firstly, working with matrix and vector representations of incidence systems had both advantages and challenges from a formalisation perspective. The main challenge lay in the considerable setup for reasoning to be transferable between the set and matrix representations. Of the theories in the final library, the incidence matrix theory by far has the most lines of code. These basic results were usually obvious, but could be tricky to prove. An initial review of the formalisation found multiple lemmas with similar statements, however the slight variations on these statements often proved useful from a proof automation standpoint in different contexts. Proof automation, specifically sledgehammer, also proved to be much less effective on matrix and vector representations than set theoretic definitions, with factors such as a much larger search space of dependencies and added assumptions on valid indexes possibly contributing to this.

Intuitive facts around linear algebraic representations, which often are what is appealing about a linear algebra proof on paper, did not always translate simply into formalisation. For example, it was common for proofs to use simple statements such as "the number of ones in an incidence vector is the size of the block it represents". In order to reason on these facts, we needed to lift a number of basic definitions of concepts already defined on lists and sets, such as counting the occurrences of a number or performing a summation over a vector. Formalising the properties of an incidence system using a matrix based definition did require extra effort in equivalence proofs, but also significantly cleaned up formal reasoning on these properties in more complex proofs at later stages. With this setup now in place, linear algebraic proofs could also potentially offer a number of benefits for formalisation. Traditional counting proofs can be very intuitive, and hard to formalise, whereas a proof of linear independence for example tends to be much more calculation based, which is easier to formalise. Additionally, linear algebraic techniques do benefit from much more significant and established libraries.

The joint focus on proof techniques and their application to formalising particular theorems presents several advantages demonstrated in this paper. The techniques in this formalisation are conceptually simple, and by providing a number of general lemmas we aim to enable formalisations using these techniques without the need to gain a deeper understanding of the complexities of the Isabelle linear algebra libraries. While there are limitations on how general these techniques can be, when applicable they provide an easy way to structure proofs of a similar nature, and reduce duplication of work, as demonstrated by their application to formalising Fisher's inequality. This application was essential in turn to refine the formal techniques, demonstrating the benefit of dually focussing on both aspects during the formalisation process. Having now established these general lemmas, it would be interesting to further experiment with applying them to other theorems in extremal combinatorics such as the well known Frankl-Wilson theorem \cite{franklIntersectionTheoremsGeometric1981}. Chapter 13 in Jukna's textbook provides inspiration for other applications of the linear algebra bound method across several different types of combinatorial structures \cite{juknaExtremalCombinatorics2011}.

\section{Conclusion}

In summary, this paper presented three primary contributions to the formalisation of combinatorics. Firstly, the formalisation of incidence matrices, providing a linear algebraic representation and basic reasoning techniques for future formal proofs on incidence systems. Secondly, we formalise general lemmas for the rank argument and linear algebra method as used in combinatorial settings, aiming to improve accessibility for utilising these techniques in a formal environment. Finally, these techniques were demonstrated through the first formalisations of a number of variations of Fisher's inequality, a consequential theorem for both set systems and the linear algebraic method in combinatorics more broadly. The final formalisation repository is available online, and will be submitted to the Isabelle AFP. 

For future work, it would be interesting to extend these techniques to more advanced settings, formalise other linear algebraic techniques and types of linear algebraic representations, and explore the potential for automation of some of the techniques involved. Lastly, Fisher's inequality itself is an important result which has several further extensions as well as applications to other interesting theorems yet to be formalised.
 
%%
%% Bibliography
%%

%% Please use bibtex, 

\bibliography{ITP22_bibv2}

\begin{thebibliography}{10}

\bibitem{aransayFormalizationExecutionLinear2014}
Jes{\'u}s Aransay and Jose Divas{\'o}n.
\newblock Formalization and {{Execution}} of {{Linear Algebra}}: {{From
  Theorems}} to {{Algorithms}}.
\newblock In Gopal Gupta and Ricardo Pe{\~n}a, editors, {\em Logic-{{Based
  Program Synthesis}} and {{Transformation}}}, volume 8901, pages 1--18.
  {Springer International Publishing}, {Cham}, 2014.

\bibitem{babaiLINEARALGEBRAMETHODS1988}
L\'{a}szl\'{o} Babai and P\'{e}ter Frankl.
\newblock {\em Linear Algebra Methods in Combinatorics}.
\newblock Department of Computer Science, University of Chicago, 2.1 edition,
  2020.
\newblock URL:
  \url{https://people.cs.uchicago.edu/~laci/CLASS/HANDOUTS-COMB/BaFrNew.pdf}.

\bibitem{ballarinLocalesLocaleExpressions2004}
Clemens Ballarin.
\newblock Locales and {{Locale Expressions}} in {{Isabelle}}/{{Isar}}.
\newblock In Stefano Berardi, Mario Coppo, and Ferruccio Damiani, editors, {\em
  Types for {{Proofs}} and {{Programs}}}, Lecture {{Notes}} in {{Computer
  Science}}, pages 34--50, {Berlin, Heidelberg}, 2004. {Springer}.

\bibitem{boseNoteFisherInequality1949}
R.~C. Bose.
\newblock A {{Note}} on {{Fisher}}'s {{Inequality}} for {{Balanced Incomplete
  Block Designs}}.
\newblock {\em The Annals of Mathematical Statistics}, 20(4):619--620, December
  1949.

\bibitem{brualdiCombinatorialMatrixTheory1991}
Richard~A. Brualdi and Herbert~J. Ryser.
\newblock {\em Combinatorial {{Matrix Theory}}}.
\newblock Encyclopedia of {{Mathematics}} and Its {{Applications}}. {Cambridge
  University Press}, {Cambridge}, 1991.

\bibitem{bukhAlgebraicMethodsCombinatoricsa}
Boris Bukh.
\newblock Lecture notes in algebraic {{Methods}} in {{Combinatorics}}: {{Rank}}
  argument, 2014.
\newblock URL: \url{http://www.borisbukh.org/AlgMethods14/rank_notes.pdf}.

\bibitem{colbournHandbookCombinatorialDesigns2007}
C.~J Colbourn and Jeffrey~H. Dinitz.
\newblock {\em Handbook of Combinatorial Designs / Edited by {{Charles J}}.
  {{Colbourn}}, {{Jeffrey H}}. {{Dinitz}}.}
\newblock Discrete Mathematics and Its Applications. {Chapman \& Hall/CRC},
  {Boca Raton, Fla. ; London}, 2nd ed. edition, 2007.

\bibitem{divasonVerifiedImplementationBerlekamp2020}
Jose Divas{\'o}n, Sebastiaan J.~C. Joosten, Ren{\'e} Thiemann, and Akihisa
  Yamada.
\newblock A {{Verified Implementation}} of the
  {{Berlekamp}}\textendash{{Zassenhaus Factorization Algorithm}}.
\newblock {\em Journal of Automated Reasoning}, 64(4):699--735, April 2020.

\bibitem{edmondsModularFirstFormalisation2021}
Chelsea Edmonds and Lawrence~C. Paulson.
\newblock A {{Modular First Formalisation}} of {{Combinatorial Design Theory}}.
\newblock In {\em {{CICM}}}, 2021.

\bibitem{fisherExaminationDifferentPossible1940a}
R.~A. Fisher.
\newblock An {{Examination}} of the {{Different Possible Solutions}} of a
  {{Problem}} in {{Incomplete Blocks}}.
\newblock {\em Annals of Eugenics}, 10(1):52--75, 1940.

\bibitem{franklIntersectionTheoremsGeometric1981}
P.~Frankl and R.~M. Wilson.
\newblock Intersection theorems with geometric consequences.
\newblock {\em Combinatorica}, 1(4):357--368, December 1981.

\bibitem{godsilToolsLinearAlgebra}
C.~D. Godsil.
\newblock Tools from {{Linear Algebra}}.
\newblock In Lov{\'a}sz~L Graham~RL, Gr{\"o}tschel~M, editor, {\em Handbook of
  {{Combinatorics}}}, volume~2. {Elsevier}, {Amsterdam}, 1996.

\bibitem{gonthierFourColourTheorem2008}
Georges Gonthier.
\newblock The {{Four Colour Theorem}}: {{Engineering}} of a {{Formal Proof}}.
\newblock In Deepak Kapur, editor, {\em Computer {{Mathematics}}}, Lecture
  {{Notes}} in {{Computer Science}}, pages 333--333, {Berlin, Heidelberg},
  2008. {Springer}.

\bibitem{gowersDimensionArgumentsCombinatorics2008}
W.~T. Gowers.
\newblock Dimension arguments in combinatorics, July 2008.
\newblock URL:
  \url{https://gowers.wordpress.com/2008/07/31/dimension-arguments-in-combinatorics/}.

\bibitem{HarrisonMatrices}
John Harrison.
\newblock The hol light theory of euclidean space.
\newblock {\em J. Autom. Reason.}, 50(2):173–190, feb 2013.
\newblock \href {https://doi.org/10.1007/s10817-012-9250-9}
  {\path{doi:10.1007/s10817-012-9250-9}}.

\bibitem{huffmanLiftingTransferModular2013}
Brian Huffman and Ond{\v r}ej Kun{\v c}ar.
\newblock Lifting and {{Transfer}}: {{A Modular Design}} for {{Quotients}} in
  {{Isabelle}}/{{HOL}}.
\newblock In David Hutchison, Takeo Kanade, Josef Kittler, Jon~M. Kleinberg,
  Friedemann Mattern, John~C. Mitchell, Moni Naor, Oscar Nierstrasz,
  C.~Pandu~Rangan, Bernhard Steffen, Madhu Sudan, Demetri Terzopoulos, Doug
  Tygar, Moshe~Y. Vardi, Gerhard Weikum, Georges Gonthier, and Michael Norrish,
  editors, {\em Certified {{Programs}} and {{Proofs}}}, volume 8307, pages
  131--146. {Springer International Publishing}, {Cham}, 2013.

\bibitem{juknaExtremalCombinatorics2011}
Stasys Jukna.
\newblock {\em Extremal {{Combinatorics}}}.
\newblock Texts in {{Theoretical Computer Science}}. {{An EATCS Series}}.
  {Springer Berlin Heidelberg}, {Berlin, Heidelberg}, 2011.

\bibitem{keinholzMatroids2018}
Jonas Keinholz.
\newblock Matroids.
\newblock {\em Isabelle Archive of Formal Proofs}, November 2018.
\newblock URL: \url{https://www.isa-afp.org/entries/Matroids.html}.

\bibitem{kroesLinearAlgebraMethods}
Daniel Kroes, Jacob Naranjo, Jiaxi Nie, Jason O'Neill, Nicholas Sieger, Sam
  Sprio, and Emily Zhu.
\newblock Lecture notes: Linear {{Algebra}} methods in {{Combinatorics}}, 2019.
\newblock URL: \url{https://mathweb.ucsd.edu/~sspiro/Abacus/AbacusF19.pdf}.

\bibitem{majumdarTheoremsCombinatoricsRelating1953}
Kulendra~N. Majumdar.
\newblock On some {{Theorems}} in {{Combinatorics Relating}} to {{Incomplete
  Block Designs}}.
\newblock {\em The Annals of Mathematical Statistics}, 24(3):377--389,
  September 1953.

\bibitem{noschinskiGraphLibraryIsabelle2015}
Lars Noschinski.
\newblock A {{Graph Library}} for {{Isabelle}}.
\newblock {\em Mathematics in Computer Science}, 9(1):23--39, March 2015.

\bibitem{obuaProvingBoundsReal2005}
Steven Obua and Technische~Universit{\"a}t M{\"u}nchen.
\newblock Proving bounds for real linear programs in isabelle/{{HOL}}.
\newblock In {\em Theorem {{Proving}} in {{Higher Order Logics}} ({{TPHOLs}}
  2005), Volume 3603 of {{Lect}}. {{Notes}} in {{Comp}}. {{Sci}}}, pages
  227--244. {Springer}, 2005.

\bibitem{soicherDesignGAPManual}
Leonard~H. Soicher.
\newblock Design {{GAP Manual}}.
\newblock URL:
  \url{https://www.gap-system.org/Manuals/pkg/design-1.7/doc/manual.pdf}.

\bibitem{stinsonCombinatorialDesignsConstructions2004}
Douglas Stinson.
\newblock {\em Combinatorial {{Designs}}: {{Constructions}} and {{Analysis}}}.
\newblock {Springer-Verlag}, {New York}, 2004.

\bibitem{thiemannFormalizingJordanNormal2016}
Ren{\'e} Thiemann and Akihisa Yamada.
\newblock Formalizing {{Jordan}} normal forms in {{Isabelle}}/{{HOL}}.
\newblock In {\em Proceedings of the 5th {{ACM SIGPLAN Conference}} on
  {{Certified Programs}} and {{Proofs}}}, {{CPP}} 2016, pages 88--99, {St.
  Petersburg, FL, USA}, January 2016. {Association for Computing Machinery}.

\bibitem{DBLP:phd/dnb/Wenzel02a}
Markus Wenzel.
\newblock {\em Isabelle, {{Isar}} - a Versatile Environment for Human Readable
  Formal Proof Documents}.
\newblock PhD thesis, Technical University Munich, Germany, 2002.
\newblock URL:
  \url{http://tumb1.biblio.tu-muenchen.de/publ/diss/in/2002/wenzel.pdf}.

\end{thebibliography}

\end{document}